\documentclass[reqno,12pt,a4paper]{article}%
\usepackage{amsfonts}
\usepackage{amsmath}
\usepackage{dpdefs}
\usepackage{lscape}%
\newtheorem{theorem}{Theorem}[section]
\newtheorem{proposition}{Proposition}[section]
\newtheorem{lemma}{Lemma}[section]
\newtheorem{assumption}{Assumption}
\hypersetup{bookmarks={true},pdfstartview={FitH}, citecolor=blue}
\begin{document}

\title{Bias Reduction of Long Memory Parameter Estimators via the Pre-filtered Sieve
Bootstrap\thanks{This research has been supported by
Australian Research Council (ARC) Discovery Grant DP120102344 and ARC Future
Fellowship FT0991045.} }
\author{D. S. Poskitt, Gael M. Martin\thanks{Corresponding author: Gael Martin,
Department of Econometrics and Business Statistics, Monash University,
Clayton, Victoria 3800, Australia. Tel.: +61-3-9905-1189; fax:
+61-3-9905-5474; email: gael.martin@monash.edu.} and Simone D. Grose\\{\small \emph{Department of Econometrics \& Business Statistics, Monash
University}}}
\maketitle

\begin{abstract}
This paper investigates the use of bootstrap-based bias correction of
semi--parametric estimators of the long memory parameter in fractionally
integrated processes. The re-sampling method involves the application of the
sieve bootstrap to data pre-filtered by a preliminary semi-parametric estimate
of the long memory parameter. Theoretical justification for using the
bootstrap techniques to bias adjust log-periodogram and semi-parametric local
Whittle estimators of the memory parameter is provided. Simulation evidence
comparing the performance of the bootstrap bias correction with analytical
bias correction techniques is also presented. The bootstrap method is shown to
produce notable bias reductions, in particular when applied to an estimator
for which analytical adjustments have already been used. The empirical
coverage of confidence intervals based on the bias-adjusted estimators is very
close to the nominal, for a reasonably large sample size, more so than for the
comparable analytically adjusted estimators. The precision of inferences (as
measured by interval length) is also greater when the bootstrap is used to
bias correct rather than analytical adjustments.

\end{abstract}

\noindent{\footnotesize \emph{MSC2010 subject classifications}: Primary 62M10,
62M15; Secondary 62G09}



\section{Introduction}

The so-called long memory, or strongly dependent, processes have come to play
an important role in time series analysis. Long range dependence, observed in
a very wide range of empirical applications, is characterized by an
autocovariance structure that decays too slowly to be absolutely summable.
Specifically, rather than the autocovariance function declining at the
exponential rate characteristic of a stable and invertible $ARMA$ process, it
declines at a hyperbolic rate dependent on a \textquotedblleft long
memory\textquotedblright\ parameter $\alpha\in(0,1)$; i.e.,
\begin{equation}
\gamma(\tau)\sim C\tau^{-\alpha},C\neq0,\text{ as }\tau\rightarrow
\infty\;.\label{H_eqn}%
\end{equation}
A detailed description of the properties of such processes can be found in
\cite{beran:1994}.

Perhaps the most popular model of a long memory process is the fractionally
integrated ($I(d)$) process introduced by \cite{granger:joyeux:1980} and
\cite{hosking:1981}. This class of processes can be characterized by the
specification,
\begin{equation}
y(t)=\sum_{j=0}^{\infty}k(j)\varepsilon(t-j)=\frac{\kappa(z)}{(1-z)^{d}%
}\,\varepsilon(t),\label{Wold}%
\end{equation}
where $\varepsilon(t)$ is zero-mean white noise, $z$ is here interpreted as
the lag operator $(z^{j}y(t)=y(t-j))$, and $\kappa(z)=\sum_{j\geq0}%
\kappa(j)z^{j}$, $\kappa(0)=1$. For any $d>-1$ the operator $(1-z)^{d}$ is
defined via the binomial expansion
\begin{equation}
(1-z)^{d}=1-dz+\frac{d(d-1)z}{2!}-\frac{d(d-1)(d-2)z^{3}}{3!}+\cdots
\,,\label{fdbinom}%
\end{equation}
and if the \textquotedblleft short memory\textquotedblright\ component
$\kappa(z)$ is the transfer function of a stable, invertible ARMA process and
$|d|<0.5$, then the coefficients of $k(z)$ are square-summable $(\sum_{j\geq
0}|k(j)|^{2}<\infty)$. In this case $y(t)$ is well-defined as the limit in
mean square of a covariance-stationary process and the model is essentially a
generalization of the classic Box-Jenkins ARIMA model \citep{box:jenkins},
\begin{equation}
(1-z)^{d}\Phi(z)y(t)=\Theta(z)\varepsilon(t),\label{eq:arfima}%
\end{equation}
in which we now allow non-integer values of the integrating parameter $d$ and
$\kappa(z)=\Theta(z)/\Phi(z).$ The long run behaviour of this process
naturally depends on the fractional integration parameter $d$. In particular,
for any $d>0$ the impulse response coefficients of the Wold representation in
\eqref{Wold} are not absolutely summable and, for $0<d<0.5$, the
autocovariances decline at the rate $\gamma(\tau)\sim C\tau^{2d-1}$ (i.e. with
reference to \eqref{H_eqn}, $\alpha=1-2d$). Such processes have been found to
exhibit dynamic behaviour very similar to that observed in many empirical time
series. See \cite{robinson:2003} for a collection of the seminal articles in
the area and \cite{doukhan:oppenheim:taqqu:2003} for a thorough review of
theory and applications. The role played by fractional processes in finance,
most notably in the modelling of the variance of financial returns, is
highlighted in \cite{andersen:bollerslev:2006} and in multiple papers
published in a special issue of \textit{Econometric Reviews }(2008, 27, Issue 1-3).

Statistical procedures for analyzing long memory processes have ranged from
the likelihood-based methods of \cite{fox:taqqu:1986}, \cite{dahlhaus:1989},
\cite{sowell:1992} and \cite{beran:1995}, to the semi-parametric techniques
advanced by \cite{geweke:porter:1983} and \cite{robinson:1995a,robinson:1995b}%
, among others. The asymptotic theory for maximum likelihood estimation (MLE)
of the parameters of such processes is well established, at least under the
assumption of Gaussian errors. In particular, we have consistency, asymptotic
efficiency, and asymptotic normality for the MLE of the fractional
differencing parameter, so providing a basis for large sample inference in the
usual manner. Such asymptotic results are, however, conditional on correct
model specification, with the MLE of $d$ typically inconsistent if either or
both the autoregressive and moving average operators in \eqref{eq:arfima} (or,
alternatively, the operator $\kappa(z)$ in \eqref{Wold}) are incorrectly
specified. The semi-parametric methods aim to produce consistent estimators of
$d$ whilst placing only very mild restrictions on the behaviour of
$\kappa(e^{\imath\lambda})$ for frequency values $\lambda$ near zero. The
semi-parametric estimators are therefore robust to different forms of short
run dynamics and offer broader applicability than a fully parametric method.
They are also asymptotically pivotal and have particularly simple asymptotic
normal distributions.

Whilst such features place the semi-parametric methods at the forefront for
use in conducting inference on $d$, the price paid for their application is a
reduction in asymptotic efficiency (relative to exact MLE) and a slower rate
of convergence to the true parameter \citep{giraitis:robinson:samarov:1997}.
Also, despite asymptotic robustness to the short run dynamics, semi-parametric
estimators have been shown to exhibit large finite sample bias in the presence
(in particular) of a substantial autoregessive component -- see
\cite{agiakloglou:newbold:wohar:1993} and \cite{lieberman:2001b} for example.
Hence, bias correction of semi-parametric estimators is an important area to explore.

In this paper we focus on bias correction of the following two semi-parametric
estimators $\widehat{d}_{T}$ of $d$:

\begin{enumerate}
\item The \cite{geweke:porter:1983}/\cite{robinson:1995a} log-periodogram
regression estimator (referred to hereafter as LPR): The ordinary least
squares (OLS) slope coefficient in a regression of $\log I_{T}(\lambda_{j})$
on a constant and $-2\log\lambda_{j}$, $j=1,\ldots,N$,
\[
\widehat{d}_{T}=\arg\min_{|d|<\half}\sum_{j=1}^{N}\left(  \log I_{T}%
(\lambda_{j})+2d(\log\lambda_{j}-\overline{\log\lambda})\right)  ^{2}\,,
\]
where $I_{T}(\lambda)=(2\pi T)^{-1}|\sum_{t=1}^{T}y(t)e^{-\imath\lambda
t}|^{2}$, the periodogram, $\lambda_{j}=2\pi j/T$, $j=1,\ldots,N$, are the
first $N$ fundamental frequencies, and $\overline{\log\lambda}=N^{-1}%
\sum_{j=1}^{N}\log\lambda_{j}$.\footnote{We have written the estimator in the
form given by \cite{robinson:1995a}. \cite{geweke:porter:1983} use the
regressor $-2\log|1-e^{-\imath\lambda_{j}}|$. The two are equivalent because
$|1-e^{-\imath\lambda}|^{2d}=|\lambda|^{2d}(1+o(1))$ as $\lambda\rightarrow
0$.}

\item The semi-parametric Gaussian (local Whittle) estimator of
\cite{robinson:1995b} (SPLW):
\[
\widehat{d}_{T}=\arg\min_{|d|<\half}\left(  \log(N^{-1}\sum_{j=1}^{N}%
\lambda_{j}^{2d}I_{T}(\lambda_{j}))-2d\overline{\log\lambda}\right)  .
\]

\end{enumerate}

Both $\widehat{d}_{T}$ are $\sqrt{N}$--CAN estimators of $d$, by which we mean
that $\widehat{d}_{T}$ is consistent ($|\widehat{d}_{T}-d|$ $=o_{p}(1)$) and
asymptotically normal, $N^{\half}(\widehat{d}_{T}-d)/\upsilon\overset
{\mathcal{D}}{\rightarrow}G(x)$, where $G(x)$ denotes the standard normal
cumulative distribution function. For the LPR estimator $\upsilon^{2}=\pi/24,$
and $\upsilon^{2}=1/4$ for the SPLW estimator. For both estimators the
bandwidth parameter $N$, denoting the number of periodogram ordinates
employed, is chosen as a monotonically increasing function of sample size $T$
, and because $\kappa(z)$ is only specified locally, $N$ must be assigned such
that $N/T\rightarrow0$ as $T\rightarrow\infty$. Too small a choice of $N$ may
prompt concern about the accuracy of the normal approximation, whereas too
large a value for $N$ entails an element of non-local averaging and is a
source of bias. In brief, although $\lim_{T\rightarrow\infty}E[\widehat{d}%
_{T}-d]=0$ the finite sample bias in such estimators can, as previously
observed, present problems.

One approach to the problem of bias is to seek an analytical solution that
will reduce the first-order bias. \cite{moulines:soulier:1999}, for example,
reduce bias by fitting a finite number of Fourier coefficients to the
logarithm of the short memory spectrum and constructing a broad-band LPR
estimator of $d$ that uses all of the frequencies in the range $(0,\pi]$, not
just those in a neighborhood of zero. \cite{andrews:guggenberger:2003}
consider a bias-adjusted estimator of $d$ obtained by including even powers of
frequency as additional regressors in the log-periodogram pseudo regression
defined in 1. above, and \cite{andrew:sun:2004} adapt this approach to the
SPLW estimator. Monte-Carlo evidence presented in
\cite{nielsen:frederiksen:2005} demonstrates the usefulness of these
bias-adjusted versions of the LPR and SPLW estimators. In particular, the
bias-corrected semi-parametric estimators are shown to outperform correctly
specified parametric estimators, although at the expense of an increase in
mean squared error.

An alternative methodological approach to bias correction, and the one that we
examine here, is to use the bootstrap. Bootstrap methodology may be thought of
as coming in two \textquotedblleft flavours\textquotedblright: the parametric,
or model-based, bootstrap, and a variety of non- or semi-parametric schemes.
The parametric bootstrap relies on having a full, correct parametric
specification for the process and is therefore at odds with the
semi--parametric approach to estimation being considered here. A less
model-dependent approach nonetheless requires a re-sampling scheme that is
able to capture the salient features of the data generating process, the
dependence structure of the process being of prime importance in the time
series context. While the block bootstrap of \cite{kunsch:1989} has
traditionally been employed for this purpose, blocking has been found to
suffer from relatively poor convergence rates. For instance, the error in the
coverage probability of a one-sided confidence interval derived from the block
bootstrap is $O(T^{-3/4})$, compared to the $O(T^{-1})$ rate achieved with
simple random samples. An attractive alternative is the \textquotedblleft
sieve\textquotedblright\ bootstrap. This works by \textquotedblleft
pre-whitening\textquotedblright\ the data using an autoregressive
approximation, with the dynamics of the process captured in a fitted
autoregression \citep[See][]{politis:2003}. Provided the order, $h$, of the
autoregression increases at a suitable rate with $T$, the convergence rates
for the sieve bootstrap are much closer (in fact arbitrarily close) to those
for simple random samples. \cite{choi:hall:2000} demonstrate the superior
convergence performance of the sieve bootstrap (over the block bootstrap) for
linear short memory processes, whilst \citet{poskitt:grose:martin:2013} build
on the results of \cite{poskitt:2008} to show that under regularity conditions
that allow for fractionally integrated $I(d)$ processes, the sieve bootstrap
achieves an error rate of $O(T^{-(1-\max\{0,d\})+\beta})$ for all $\beta>0$,
for a general class of statistics.

The current paper uses a modified sieve bootstrap to bias-correct the LPR and
SPLW estimators of the memory parameter in fractionally integrated $I(d)$
processes. The bootstrap method uses a consistent semi-parametric estimator of
the long memory parameter to pre-filter the raw data, \textit{prior to} the
use of a long autoregressive approximation as the `sieve' from which bootstrap
samples are produced.
The bias correction proceeds in an iterative fashion, with a stochastic
stopping rule invoked to produce the final estimator.
Starting with $\sqrt{N}$--CAN estimators that satisfy a requisite Edgeworth
expansion and large-deviation properties we derive error rates for estimating
the bias of both the LPR and SPLW estimators, with the accuracy with which the
bootstrap method estimates the bias in finite samples then documented in a
simulation setting. We also use the bootstrap method to bias-adjust the
(already) analytically bias-adjusted versions of these two estimators. The
analytically bias-adjusted LPR estimator of \cite{andrews:guggenberger:2003}
(LPR-BA hereafter) is produced as the OLS coefficient of the regressor
$-2\log\lambda_{j}$ in the regression of $\log I_{T}(\lambda_{j})$ on a
constant, $-2\log\lambda_{j}$, and $\lambda_{j}^{2p}$, $p=1,\ldots,P$,
$j=1,\ldots,N$. The analytically bias-adjusted SPLW estimator of
\cite{andrew:sun:2004} (SPLW-BA hereafter) is produced as the first element of
$(\widehat{d}_{T},\widehat{\theta}_{1},\ldots,\widehat{\theta}_{P})=\arg\min
LW(d,\theta_{1},\ldots,\theta_{P})$ where
\begin{align*}
LW(d,\theta_{1},\ldots,\theta_{P}) &  =\log\left(  N^{-1}\sum_{j=1}^{N}%
\lambda_{j}^{2d}I_{T}(\lambda_{j})\exp\left\{  \sum_{p=1}^{P}\theta_{p}%
\lambda_{j}^{2p}\right\}  \right) \\
&  -N^{-1}\sum_{j=1}^{N}\left\{  \sum_{p=1}^{P}\theta_{p}\lambda_{j}%
^{2p}\right\}  -2d\overline{\log\lambda}\,.
\end{align*}

The paper proceeds as follows. Section 2 briefly summarizes the statistical
properties of long memory processes, and outlines the sieve bootstrap (both
`raw' and pre-filtered) in this context. The pre-filtered sieve bootstrap bias
adjustment (PFSA(BA)) algorithm is also described in this section. In Section
3 we present the relevant approximations and exploit these to produce the
error rates for the bootstrap technique. Section 4 outlines the iterated
version of the bootstrap bias correction technique. Details of the simulation
study are given in Section 5. Section 6 closes the paper.

\section{Long memory processes, autoregressive approximation, and the sieve
bootstrap}

Let $y(t)$ for $t\in{\mathcal{Z}}$ denote a linearly regular,
covariance-stationary process, with representation as in \eqref{Wold} where;

\begin{assumption}
\label{Ass1} The transfer function in the representation \eqref{Wold} is given
by $k(z)=\kappa(z)/(1-z)^{d}$ where $|d|<0.5$ and $\kappa(z)\neq0$, $|z|\leq
1$. The impulse response coefficients of $\kappa(z)$ satisfy $k(0)=1$ and
$\sum_{j\geq0}j|\kappa(j)|<\infty$.
\end{assumption}

\begin{assumption}
\label{Ass2} The innovations process $\varepsilon(t)$ is an i.i.d. zero mean
Gaussian white noise process with variance $\sigma^{2}$.
\end{assumption}

Assumption \ref{Ass1} serves to characterize the spectral features of quite a
wide class of processes, including the ARFIMA family of models that are the
focus of this paper. This assumption implies that the innovations in the
unilateral representation in \eqref{Wold} are \textit{fundamental}; meaning
that $\varepsilon(t)$ lies in the space spanned by current and past values of
$y(t)$, and $\varepsilon(t)$ and $y(s)$ are uncorrelated for all $s<t$. For a
discussion of the role of fundamentalness in the context of the autoregressive
sieve bootstrap see \citet{kreiss:paparoditis:politis:2011}. Note that the
regularity conditions employed in \citet{kreiss:paparoditis:politis:2011}
exclude fractional time series, but using the extension of Baxter's inequality
to long range dependent processes due to \citet{inoue:kasahara:2006} it is
possible to generalize the results of \citet{kreiss:paparoditis:politis:2011}
to time series generated from a fractional transformation of a linear
processes. In particular, since the statistics that we investigate are
asymptotically pivotal the results in
\citet[][Section 3]{kreiss:paparoditis:politis:2011} can be extended to the
statistics and class of processes under consideration here.

Assumption \ref{Ass2} implies that $y(t)$ is a Gaussian process. A basic
property of such a process that underlies the sieve bootstrap methodology and
the associated results is that $y(t)$ is linearly regular and the linear
predictor
\begin{equation}
\bar{y}(t)=\sum_{j=1}^{\infty}\pi(j)y(t-j)\,,\label{plin}%
\end{equation}
where $\sum_{j=0}^{\infty}\pi(j)z^{j}=(1-z)^{d}\kappa(z)^{-1}$, is the minimum
mean squared error (MMSE) predictor (MMSEP) of $y(t)$ based upon its entire
past. The need to invoke Gaussianity is unfortunate but is unavoidable here as
we wish employ certain results from the existing literature where the
assumption that $y(t)$ is a Gaussian process is adopted. It is likely that our
results can be extended to more general linear processes, although the
regularity conditions and prerequisites needed for such extensions are liable
to be relatively involved.

The MMSEP of $y(t)$ based only on a finite number (h) of past observations
(MMSEP(h)) is then
\begin{equation}
\bar{y}_{h}(t)=\sum_{j=1}^{h}\pi_{h}(j)y(t-j)\equiv-\sum_{j=1}^{h}\phi
_{h}(j)y(t-j);\label{plinh}%
\end{equation}
where the minor reparameterization from $\pi_{h}$ to $\phi_{h}$ allows us, on
also defining $\phi_{h}(0)=1$, to conveniently write the corresponding
prediction error as
\begin{equation}
\varepsilon_{h}(t)=\sum_{j=0}^{h}\phi_{h}(j)y(t-j).\label{epsh}%
\end{equation}
The finite-order autoregressive coefficients $\phi_{h}(1),\ldots,\phi_{h}(h)$
can be deduced from the Yule-Walker equations
\begin{equation}
\sum_{j=0}^{h}\phi_{h}(j)\gamma(j-k)=\delta_{0}(k)\sigma_{h}^{2}\,,\quad
k=0,1,\ldots,h,\label{Y-W}%
\end{equation}
in which $\gamma(\tau)=\gamma(-\tau)=E[y(t)y(t-\tau)]$, $\tau=0,1,\ldots,$ is
the autocovariance function of the process $y(t)$, $\delta_{0}(k)$ is
Kronecker's delta (i.e., $\delta_{0}(k)=0\; \forall\;k\neq0;\; \delta
_{0}(0)=1$), and the MMSE is
\begin{equation}
\sigma_{h}^{2}=E\big[\varepsilon_{h}(t)^{2}\big]\,,\label{Varh}%
\end{equation}
the prediction error variance associated with $\bar{y}_{h}(t)$.

The use of finite-order AR models to approximate an unknown (but suitably
regular) process therefore requires that the optimal predictor $\bar{y}%
_{h}(t)$ determined from the autoregressive model of order $h$ be a good
approximation to the \textquotedblleft infinite-order\textquotedblright%
\ predictor $\bar{y}(t)$ for sufficiently large $h$. The asymptotic validity,
and properties, of finite-order autoregressive models when $h\rightarrow
\infty$ with the sample size $T$ under regularity conditions that admit
non-summable processes was proved in \cite{poskitt:2007}. Briefly, the
order-$h$ prediction error $\varepsilon_{h}(t)$ converges to $\varepsilon(t)$
in mean-square, the estimated sample-based covariances converge to their
population counterparts\emph{\ }(though at a slower rate than for a
conventional $I(0)$ stationary process) and the ordinary least squares (\LS)
and \YW\ estimators of the coefficients of the approximating autoregression
are asymptotically equivalent and consistent. Furthermore, order selection by
AIC, a commonly employed practice in the context of the sieve bootstrap
\citep[][\S 3]{politis:2003}, is asymptotically efficient in the sense of
minimizing Shibata's \citeyearpar{shibata:1980} figure of merit. The sieve
bootstrap, with order selected via an asymptotically efficient criterion, is
accordingly a plausible \textquotedblleft non-parametric\textquotedblright%
\ bootstrap technique for long memory processes.

\subsection{The raw sieve bootstrap}

Details of the raw sieve bootstrap (SB) for fractional processes are given in
\cite{poskitt:2008}. For convenience we reproduce here the basic steps of the
SB algorithm for generating a realization of a process $y(t)$, prior to
presenting the PFSB(BA) algorithm adopted for bias-adjustment in this paper.

\bigskip

The raw sieve bootstrap (SB) algorithm:

\begin{description}
\item[SB1.] \label{s1} Given data $y(t)$, $t=1,\ldots,T$, and using
$y(1-j)=y(T-j+1)$, $j=1,\ldots,h$, as initial values, calculate parameter
estimates of the $AR(h)$ approximation, denoted by $\bar{\phi}_{h}%
(1),\ldots,\bar{\phi}_{h}(h)$, and evaluate the residuals
\[
\bar{\varepsilon}_{h}(t)=\sum_{j=0}^{h}\bar{\phi}_{h}(j)y(t-j)\,,\,t=1,\ldots
,T\,,
\]
From $\bar{\varepsilon}_{h}(t)$, $t=1,\ldots,T$, construct the standardized
residuals $\tilde{\varepsilon}_{h}(t)=(\bar{\varepsilon}_{h}(t)-\bar
{\varepsilon}_{h})/\bar{\sigma}_{h}$ where $\bar{\varepsilon}_{h}=T^{-1}%
\sum_{t=1}^{T}\bar{\varepsilon}_{h}(t)$ and $\bar{\sigma}_{h}^{2}=T^{-1}%
\sum_{t=1}^{T}(\bar{\varepsilon}_{h}(t)-\bar{\varepsilon}_{h})^{2}$.

\item[SB2.] Let $\varepsilon_{h}^{+}(t)$, $t=1,\ldots,T$, denote a simple
random sample of \textit{i.i.d.} values drawn from
\[
U_{\bar{\varepsilon}_{h},T}(e)=T^{-1}\sum_{t=1}^{T}\mathbf{1}\{\tilde
{\varepsilon}_{h}(t)\leq e\}\,,
\]
the probability distribution function that places a probability mass of $1/T$
at each of $\tilde{\varepsilon}_{h}(t)$, $t=1,\ldots,T$. Set $\varepsilon
_{h}^{\ast}(t)=\bar{\sigma}_{h}\varepsilon_{h}^{+}(t)$, $t=1,\ldots,T$.

\item[SB3.] Construct the sieve bootstrap realization $y^{\ast}(1),\ldots
,y^{\ast}(T)$, where $y^{\ast}(t)$ is generated from the autoregressive
process
\[
\sum_{j=0}^{h}\bar{\phi}_{h}(j)y^{\ast}(t-j)=\varepsilon_{h}^{\ast
}(t)\,,\,t=1,\ldots,T\,,
\]
initiated at $y^{\ast}(1-j)=y(\tau-j+1)$, $j=1,\ldots,h$, where $\tau$ is a
discrete uniform random variable with support on the integers $h,\ldots,T$.
\end{description}

Crucially, the rate of convergence of the coefficient estimates $\bar{\phi
}_{h}(1),\ldots,\bar{\phi}_{h}(h)$ evaluated in Step SB1 is dependent upon the
value of the fractional index $d$, as formalized in the following theorem

\begin{theorem}
\label{consistentYW} Let $\sum_{j=0}^{h}\bar{\phi}_{h}(j)z^{j}$ denote the
Burg, \LS\ or \YW\ estimator of $\sum_{j=0}^{h}\phi_{h}(j)z^{j}$. If $y(t)$ is
a linearly regular, covariance-stationary process that satisfies Assumptions
\ref{Ass1} and \ref{Ass2}, then for all $h\leq H_{T}=a(\log T)^{c}$, $a>0$,
$c<\infty$,
\[
\sum_{j=1}^{h}|\bar{\phi}_{h}(j)-\phi_{h}(j)|^{2}=O\left(  h\left(  \frac{\log
T}{T}\right)  ^{1-2\,max\{0,d\}}\right)
\]
with probability one.
\end{theorem}

The proof of this theorem is placed in Appendix A, along with the proofs of
other results presented in the paper.

\subsection{The pre-filtered sieve bootstrap}

Given the dependence of the convergence of $\sum_{j=0}^{h}\bar{\phi}%
_{h}(j)z^{j}$ to $\sum_{j=0}^{h}\phi_{h}(j)z^{j}$ on the value of $d$, the
convergence of any bootstrap generated sampling distribution to the true
unknown sampling distribution is also dependent on the value of $d$, see
\cite{poskitt:2008}. In particular, in \citet{poskitt:grose:martin:2013} it is
shown that under appropriate regularity the raw sieve bootstrap achieves a
convergence rate of $O(T^{-(1-\max\{0,d\})+\beta})$ for all $\beta>0$.
Obviously, in the long memory case where $0<d<0.5$, the closer is $d$ to zero
the closer the convergence rate of $O(T^{-(1-d)+\beta})$ will be to the rate
$O(T^{-1+\beta})$ achieved with short memory (and anti-persistent) processes.
The empirical regularity of estimated values of $d$ in the $0<d<0.5$ range
thus provides motivation for the idea of pre-filtering the series prior to the
application of the sieve. Specifically, we employ a modified sieve method
wherein, for a given preliminary value of $d$, we pre-filter the data using
this value, apply the AR approximation (and sieve bootstrap) to the
pre-filtered data, before using the inverse filter to produce the final
realization of $y(t).$ With this procedure, the raw sieve is applied (by
construction) to filtered data with shorter memory; hence the achievement of
an improved convergence rate.

For any $d>-1$ let $(1-z)^{d}=\sum_{j=0}^{\infty}\alpha_{j}^{(d)}z^{j}$ where
$\alpha_{j}^{(d)}$, $j=0,1,2,\ldots$, denote the coefficients of the
fractional difference operator when expressed in terms of its binomial
expansion, as in the right hand side of \ref{fdbinom}. Given a preliminary
value $d^{f}$ of $d$, pre-filtered sieve bootstrap (PFSB) realizations of
$y(t)$ are generated using the following algorithm:

\begin{description}
\item[PFBS1.] Calculate the coefficients of the filter $(1-z)^{d^{f}}$ and
from the data generate the filtered values
\[
w^{f}(t)=\sum_{j=0}^{t-1}\alpha_{j}^{(d^{f})}y(t-j)
\]
for $t=1,\ldots,T$.

\item[PFBS2.] Fit an AR approximation to $w^{f}(t)$ and generate a sieve
bootstrap sample $w^{\ast_{f}}(t)$, $t=1,\ldots,T$, of the filtered data as in
Steps SB1--SB3 of the SB algorithm, with $w^{f}(t)$ and $w^{\ast_{f}}(t)$
playing the role of $y(t)$ and $y^{\ast}(t)$ respectively therein.

\item[PFBS3.] Using the coefficients of the (inverse) filter $(1-z)^{-d^{f}}$
construct a corresponding pre-filtered sieve bootstrap draw
\[
y^{\ast_{f}}(t)=\sum_{j=0}^{t-1}\alpha_{j}^{(-d^{f})}w^{\ast_{f}}(t-j)
\]
of $y(t)$ for $t=1,\ldots,T$, where the superscript $f$\ is used to
distinguish this bootstrap draw from the bootstrap draw produced by the raw
sieve algorithm, in Step SB3 above.
\end{description}

In \citet{poskitt:grose:martin:2013} it is shown that given a judicious choice
of $d^{f}$ shorter memory will be induced by the preliminary filtering at Step
PFSB1. The accuracy of the AR approximation and, thereby, the sieve bootstrap
in Step PFSB2 will accordingly be increased, and this increase in accuracy
will be passed on to the PFSB draws in Step PFSB3, resulting in a convergence
rate equal to that obtained in the short memory case, namely $O(T^{-1+\beta}%
)$. Using these results as motivation we proceed to work with the PFSB
algorithm for the purpose of bias adjustment. More formal theoretical
justification of the validity of the pre-filtered sieve when used for this
particular purpose is provided in Section 3.

\subsection{Bias correction via the pre-filtered sieve bootstrap}

To bias adjust a chosen estimator, $\widehat{d}_{T}$, of $d$ we proceed as follows:

\begin{description}
\item[BA1.] Calculate $\widehat{d}_{T}$ from the data $y(t)$, $t=1,\ldots,T$.

\item[BA2.] Use $d^{f}$ as the preliminary value in Steps PFSB1-PFSB3 of the
PFSB algorithm and produce $B$ bootstrap realizations $y_{b}^{\ast_{f}}(t)$,
$t=1,\ldots,T$, $b=1,\ldots,B$, of the process $y(t)$. From these construct
$B$ bootstrap values of the estimator, $\widehat{d}_{T,b}^{\ast_{f}}$,
$b=1,2,...,B$, by evaluating the estimator $\widehat{d}_{T}$ for each of the
$B$ independent bootstrap draws.

\item[BA3.] Estimate the bias of $\widehat{d}_{T}$ by
\begin{equation}
\widehat{b}_{T,B}^{\ast_{f}}=\left(  \frac{1}{B}%
{\textstyle\sum\limits_{b=1}^{B}}
\widehat{d}_{T,b}^{\ast_{f}}\right)  -d^{f}\label{bs_bias}%
\end{equation}
and produce the bias-adjusted estimator
\begin{equation}
\widetilde{d}_{T}=\widehat{d}_{T}-\widehat{b}_{T,B}^{\ast_{f}}%
.\label{bias_adjusted_est}%
\end{equation}

\end{description}

We refer to this as the PFSB(BA) algorithm.

\section{Some Theoretical Underpinnings}

The use of the PFSB(BA) algorithm to correct the finite sample bias of an
estimator is justified only if the method produces a bootstrap distribution
that copies the true sampling distribution of the estimator to the appropriate
order of magnitude. Not surprisingly, the rate of convergence of the bootstrap
to the true sampling distribution is shown to be dependent on the proximity of
the preliminary value employed in the PFSB, namely $d^{f}$, to the true value
of $d$, as well as the order ($h)$ of the autoregressive approximation used in
the sieve component of the PFSB. Presuming that $d^{f}$ is itself estimated
from the data, $d^{f}=d_{T}^{f}$ say, the main content of these findings are
presented in Theorems \ref{Biasadj} and \ref{pfsbbias}.

To begin, suppose that $\widehat{d}_{T}$ (the estimator to be bias-corrected)
is an asymptotically pivotal $\sqrt{N}$--CAN estimator of $d$ and that the
sampling distribution of $N^{\half}(\widehat{d}_{T}-d)$ admits an Edgeworth
expansion such that
\begin{equation}
\sup_{x}\left\vert \textmd{P}\{\frac{N^{\half}(\widehat{d}_{T}-d)}{\upsilon
}<x\}-G(x)\right\vert =o\left(  \frac{N^{5/2}}{T^{2}}\right) \label{dEdge}%
\end{equation}
where $G(\cdot)$ denotes the standard normal distribution function. Let
$b_{T}$ denote the finite sample bias of $\widehat{d}_{T}$, that is,
\begin{equation}
b_{T}=E[\widehat{d}_{T}]-d.\label{bt}%
\end{equation}
Since $\lim_{T\rightarrow\infty}N^{\half}E[\widehat{d}_{T}-d]=0$ we have
$b_{T}=o(N^{-\half})$ (recall that $N\rightarrow\infty$ as $T\rightarrow
\infty$ such that $N/T\rightarrow0$). Here $E$ denotes expectation taken with
respect to the original probability space $(\Omega,\mathfrak{F},P)$.
Substituting \eqref{bt} into \eqref{dEdge} gives the approximation
\begin{align}
\textmd{P}\{N^{\half}(\widehat{d}_{T}-E[\widehat{d}_{T}])<x\}  &
=\textmd{P}\{N^{\half}(\widehat{d}_{T}-d)<x+b_{T}\}\nonumber\label{Chib}\\
& =G((x+N^{\half}b_{T})/\upsilon)+o(N^{-\half})
\end{align}
for the distribution of the finite sample deviation $\widehat{d}%
_{T}-E[\widehat{d}_{T}]$.

Now let $\widehat{d}_{T}^{\ast_{f}}$ denote the value of $\widehat{d}_{T}$
calculated from a bootstrap realization of the process, $y^{\ast_{f}}(t)$,
$t=1,\ldots,T$, constructed using the PFSB algorithm where; (i) the
pre-filtering value $d_{T}^{f}$ satisfies the conditions stated above for
$\widehat{d}_{T}$ and, for the sake of argument; (ii) the innovations
$\varepsilon_{h}^{\ast}(t)$, $t=1,\ldots,T$, used in Step PFSB2 are generated
as i.i.d. $N(0,\bar{\sigma}_{h}^{2})$. Since the process $\varepsilon
_{h}^{\ast}(t)$ is now explicitly Gaussian, it follows that $y^{\ast_{f}}(t)$
will be a fractionally integrated $AR(h)$ Gaussian process with parameters
$d_{T}^{f}$ and $\bar{\phi}_{h}(1),\ldots,\bar{\phi}_{h}(h)$, and
\begin{equation}
\sup_{x}\left\vert \textmd{P}^{\ast}\{\frac{N^{\half}(\widehat{d}_{T}%
^{\ast_{f}}-d_{T}^{f})}{\upsilon}<x\}-G(x)\right\vert =o\left(  \frac{N^{5/2}%
}{T^{2}}\right) \label{dEdge*}%
\end{equation}
where $(\Omega^{\ast},\mathfrak{F}^{\ast},P^{\ast})$ denotes the probability
space induced by the bootstrap process.\footnote{The innovations generated in
Step PFBS2 are i.i.d. $(0,\bar{\sigma}_{h}^{2})$ by construction (see Steps
SB1--SB2 of the SB algorithm), and when $y(t)$ is Gaussian we can expect
$\varepsilon_{h}^{\ast}(t)$, $t=1,\ldots,T$, based upon Steps SB1--SB2 to be
approximately Gaussian. Replacing the innovations generated in Step PFSB2 by
i.i.d. $N(0,\bar{\sigma}_{h}^{2})$ innovations in the simulations (as would be
strictly necessary to accord with the theoretical derivations) produced
results that were virtually indistinguishable from those reported in Section 5
below.} Denote the expectation associated with $(\Omega^{\ast},\mathfrak{F}%
^{\ast},P^{\ast}) $ by $E^{\ast}$. Proceeding as previously, replacing
$\widehat{d}_{T}$ by $\widehat{d}_{T}^{\ast_{f}}$, $d$ by $d_{T}^{f}$ and
$E[\widehat{d}_{T}]$ by $E^{\ast}[\widehat{d}_{T}^{\ast_{f}}]=d_{T}^{f}%
+b_{T}^{\ast}$, with
\begin{equation}
b_{T}^{\ast}=E^{\ast}[\widehat{d}_{T}^{\ast_{f}}]-d_{T}^{f}\label{btbar}%
\end{equation}
by construction, and substiting \eqref{btbar} into \eqref{dEdge*} we obtain
the approximation
\begin{align}
\textmd{P}^{\ast}\{N^{\half}(\widehat{d}_{T}^{\ast_{f}}-E^{\ast}[\widehat
{d}_{T}^{\ast_{f}}])<x\}  & =\textmd{P}^{\ast}\{N^{\half}(\widehat{d}%
_{T}^{\ast_{f}}-d_{T}^{f})<x+b_{T}^{\ast}\}\nonumber\label{Chib*}\\
& =G((x+N^{\half}b_{T}^{\ast})/\upsilon)+o(N^{-\half})
\end{align}
for the bootstrap deviation $\widehat{d}_{T}^{\ast_{f}}-E^{\ast}[\widehat
{d}_{T}^{\ast_{f}}]$.

For a discussion of consistency and asymptotic normality of the LPR and SPLW
estimators
see, for example, \citet{hurvich:deo:brodsky:1998} and
\citet{giraitis:robinson:2003} respectively. \citet{giraitis:robinson:2003}
also present Edgeworth expansions for the SPLW estimator.
\citet{lieberman:rousseau:zucker:2001} develop Edgeworth expansions for
quadratic forms in Gaussian long memory series, and
\cite{fay:moulines:soulier:2004} provide a discussion of Edgeworth expansions
in the context of linear statistics applied to long range dependent linear
processes, with extensions to the LPR estimator presented in \citet{fay:2010}.
From these references we can glean that the preceding $\sqrt{N}$--CAN and
Edgeworth requisites require that the bandwidth parameter $N$ be chosen such
that $N\sim KT^{\nu}$ where $2/3<\nu<4/5$, $K\in(0,\infty)$.
Asymptotic normality of the estimators considered here requires that
$N=o(T^{4/5})$, hence the upper bound on $\nu$. The lower bound on $\nu$
reflects that unless $N$ increases sufficiently quickly with $T$ terms due to
bias of order $O(\log^{3}N/N)$ (see \eqref{biasd} and \eqref{biasbsd} below)
compete with more standard terms in the Edgeworth expansions.

\begin{theorem}
\label{Biasadj} Suppose that the process $y(t)$ satisfies Assumptions
\ref{Ass1} and \ref{Ass2}, and that the PFSB algorithm is applied to
$\widehat{d}_{T}$ using the preliminary value $d_{T}^{f}$ and an AR$(h)$
approximation. Assume that $d_{T}^{f}$ and $\widehat{d}_{T}$ are $\sqrt{N}%
$--CAN estimators with bandwidth parameter chosen such that $N\sim KT^{\nu}$
where $2/3<\nu<4/5$, $K\in(0,\infty)$. Then for all $h\leq H_{T}=a(\log
T)^{c}$, $a>0$, $c<\infty$,
\begin{align*}
&  \sup_{x}\left\vert \textmd{P}\{N^{\half}(\widehat{d}_{T}-E[\widehat{d}%
_{T}])<x\}-\textmd{P}^{\ast}\{N^{\half}(\widehat{d}_{T}^{\ast_{f}}-E^{\ast
}[\widehat{d}_{T}^{\ast_{f}}])<x\}\right\vert \\
&  =O(N^{\half}\left\vert b_{T}-b_{T}^{\ast}\right\vert )+o(N^{-1/2})\,,
\end{align*}
%
where $b_{T}$ and $b_{T}^{\ast}$ are as defined in \eqref{bt} and
\eqref{btbar} respectively.
\end{theorem}

Theorem \ref{Biasadj} indicates that if the bandwidth of the estimators
$d_{T}^{f}$ and $\widehat{d}_{T}$ is chosen appropriately then the PFSB
distribution of $N^{\half}(\widehat{d}_{T}^{\ast_{f}}-E^{*}[\widehat{d}%
^{*_{f}}_{T}])$ will closely approximate the true finite sampling distribution
of $N^{\half}(\widehat{d}_{T}-E[\widehat{d}_{T}])$ provided $N^{\half}%
|b_{T}-\bar{b}_{T}|$ is sufficiently small. Given that $\widehat{b}%
_{T,B}^{\ast_{f}}$ in \eqref{bs_bias} can be made arbitrarily close to the
finite sample bias induced by the PFSB distribution by taking $B$ sufficiently
large, we can therefore anticipate that if $N^{\half}|b_{T}-\bar{b}%
_{T}|\rightarrow0$ sufficiently quickly then $N^{\half}(\widehat{d}_{T}%
^{\ast_{f}}-E^{*}[\widehat{d}^{*_{f}}_{T}])$, with the sample mean of $B$
bootstrap draws used to represent $E^{\ast}[\widehat{d}_{T}^{\ast_{f}}]$, will
closely approximate $N^{\half}(\widehat{d}_{T}-E[\widehat{d}_{T}])$. This then
provides a justification for using the PFSB(BA) algorithm to estimate the bias
of $\widehat{d}_{T}$ and, in turn, produce the bias-adjusted estimate.

To evaluate the magnitude of $|b_{T}-b_{T}^{\ast}|$ note that $|\kappa
(e^{\imath\lambda})|^{2}$ is a bounded, even function of $\lambda$, and we
have the power series (McLaurin) expansion $|\kappa(e^{\imath\lambda}%
)|^{2}=c_{0}+\sum_{j\geq1}c_{j}|\lambda|^{2j}=c_{0}+c_{1}|\lambda
|^{2}+o(|\lambda|^{3})$ as $|\lambda|\rightarrow0$. Then it can be shown that
\begin{equation}
b_{T}=-\beta\frac{2c_{1}}{9c_{0}}\left(  \frac{N}{T}\right)  ^{2}+o\left(
\frac{N^{2}}{T^{2}}\right)  +O\left(  \frac{\log^{3}N}{N}\right)
,\label{biasd}%
\end{equation}
where $\beta=1/(4\pi^{2})$ for the SPLW estimator
\citep{giraitis:robinson:2003} and $\beta=\pi^{2}$ for the LPR estimator
\citep{hurvich:deo:brodsky:1998}. Similarly, set $\bar{\kappa}_{h}%
(z)=\sum_{j=0}^{\infty}\bar{\kappa}_{h}(j)z^{j}$ where the $\bar{\kappa}%
_{h}(j)$ and $\bar{\phi}_{h}(j)$ are related by the recursions
\begin{equation}
\bar{\phi}_{h}(0)=\bar{\kappa}_{h}(0)=1\,,~~\sum_{i=0}^{j}\bar{\kappa}%
_{h}(i)\bar{\phi}_{h}(j-i)=0,\;j=1,2,\ldots\,.\label{phihinv}%
\end{equation}
By construction $\bar{\kappa}_{h}(z)\bar{\phi}_{h}(z)=1$ for all $|z|\leq1$
and $\bar{\kappa}_{h}(z)$ yields the AR$(h)$ approximation to $\kappa(z)$
implicit in the PFSB. Then $|\bar{\kappa}_{h}(e^{\imath\lambda})|^{2}%
=|\sum_{j=0}^{h}\bar{\phi}_{h}(j)e^{\imath\lambda j}|^{-2}=\bar{c}_{0}+\bar
{c}_{1}|\lambda|^{2}+o(|\lambda|^{3})$ as $|\lambda|\rightarrow0 $ and
\begin{equation}
b_{T}^{\ast}=-\beta\frac{2\bar{c}_{1}}{9\bar{c}_{0}}\left(  \frac{N}%
{T}\right)  ^{2}+o\left(  \frac{N^{2}}{T^{2}}\right)  +O\left(  \frac{\log
^{3}N}{N}\right)  \,.\label{biasbsd}%
\end{equation}
Simple algebraic manipulation applied to \ref{biasd} and \ref{biasbsd} gives
us the following bound
\begin{align*}
|b_{T}-b_{T}^{\ast}|  & =\beta\frac{2}{9}\left\vert \frac{\bar{c}_{1}}{\bar
{c}_{0}}-\frac{c_{1}}{c_{0}}\right\vert \left(  \frac{N^{2}}{T^{2}}\right)
+o\left(  \frac{N^{2}}{T^{2}}\right)  +O\left(  \frac{\log^{3}N}{N}\right) \\
& \leq\beta\frac{2}{9}\left(  \left\vert \frac{c_{1}(\bar{c}_{0}-c_{0})}%
{c_{0}\bar{c}_{0}}\right\vert +\left\vert \frac{(\bar{c}_{1}-c_{1})}{\bar
{c}_{0}}\right\vert \right)  \left(  \frac{N^{2}}{T^{2}}\right)  +o\left(
\frac{N^{2}}{T^{2}}\right)  +O\left(  \frac{\log^{3}N}{N}\right)  \,.
\end{align*}
The magnitude of $|b_{T}-b_{T}^{\ast}|$ obviously depends on the order of
$(\bar{c}_{0}-c_{0})$ and $(\bar{c}_{1}-c_{1})$, and note that larger
bandwidth entails larger bias and the need for more precise correction via the
AR$(h)$ approximation to the short memory spectrum.

Let $\phi_{h}^{f}(z)=\sum_{j=0}^{h}\phi_{h}^{f}(j)z^{j}$ where $\phi_{h}%
^{f}(1),\ldots,\phi_{h}^{f}(h)$ denote the coefficients in the MMSEP(h) of the
process
\[
w^{f}(t)=(1-z)^{d^{f}}y(t)=\frac{\kappa(z)}{(1-z)^{d-d^{f}}}\,\varepsilon
(t)\,,
\]
and let $\sigma_{h}^{f2}$ denote the MMSE. Set $\kappa^{f}(z)=\kappa
(z)/(1-z)^{d-d^{f}}$ and define $\kappa_{h}^{f}(z)=\{\phi_{h}^{f}(z)\}^{-1} $
by replacing the coefficients of $\bar{\phi}_{h}(z)$ by those of $\phi_{h}%
^{f}(z)$ in the recursions in equation \eqref{phihinv}. The magnitude of
$(\bar{c}_{0}-c_{0})$ and $(\bar{c}_{1}-c_{1})$ can now be derived from the
following lemma.

\begin{lemma}
\label{kk} Suppose that the process $y(t)$ satisfies Assumptions \ref{Ass1}
and \ref{Ass2}, and that the PFSB algorithm is applied using; a preliminary
value $d^{f}=d_{T}^{f}$ where $d_{T}^{f}$ is such that $|d_{T}^{f}%
-d|<\delta_{T}$ where $\delta_{T}\log T\rightarrow0$ almost surely ($a.s.$) as
$T\rightarrow\infty$, and an AR$(h)$ approximation where $h\leq H_{T}=a(\log
T)^{c}$, $a>0$, $c<\infty$. Then
\[
\lim_{T\rightarrow\infty}\left\vert |\bar{\kappa}_{h}(e^{\imath\lambda}%
)|^{2}-|\kappa(e^{\imath\lambda})|^{2}\right\vert \leq\nu_{1,T}+\nu_{2,T}%
+\nu_{3,T}%
\]
where for all $\lambda\in\lbrack2\pi/T,2\pi N/T]$
\begin{align*}
\nu_{1,T}  & =\left\vert |\bar{\kappa}_{h}(e^{\imath\lambda})|^{2}-|\kappa
_{h}^{f}(e^{\imath\lambda})|^{2}\right\vert =O(h(\log T/T)^{\half-\delta_{T}%
})\\
\nu_{2,T}  & =\left\vert |\kappa_{h}^{f}(e^{\imath\lambda})|^{2}-|\kappa
^{f}(e^{\imath\lambda})|^{2}\right\vert =O(\delta_{T}h^{-|d|})\quad
\mbox{and}\\
\nu_{3,T}  & =\left\vert |\kappa^{f}(e^{\imath\lambda})|^{2}-|\kappa
(e^{\imath\lambda})|^{2}\right\vert =O(\delta_{T}\log T)\,.
\end{align*}
with probability one.
\end{lemma}

Lemma \ref{kk} leads to the following result.

\begin{theorem}
\label{pfsbbias} Suppose that the conditions in Theorem \ref{Biasadj} hold.
Assume also that $b_{T}=E[\widehat{d}_{T}]-d$ and $\overline{b}_{T}%
=E[\widehat{d}_{T}^{\ast_{f}}]-d_{T}^{f}$ are expressed as in \ref{biasd} and
\ref{biasbsd} respectively. If the PFSB algorithm is applied using; a
preliminary value $d^{f}=d_{T}^{f}$ where $d_{T}^{f}$ is such that $|d_{T}%
^{f}-d|<\delta_{T}$ where $\delta_{T}\log T\rightarrow0$ $a.s.$ as
$T\rightarrow\infty$, and an AR$(h)$ approximation where $h\leq H_{T}=a(\log
T)^{c}$, $a>0$, $c<\infty$, then
\[
|b_{T}-b_{T}^{\ast}|=O\left(  \max\{h(\frac{\log T}{T})^{\half-\delta_{T}%
},\delta_{T}h^{-|d|},\delta_{T}\log T\}\right)  +o\left(  \frac{N^{2}}{T^{2}%
}\right)  \quad a.s.
\]

\end{theorem}

As preempted above, the convergence of $\bar{b}_{T}$ to $b_{T}$ depends on the
order of the autoregressive approximation and the proximity of the preliminary
value employed in the PFSB to the true $d$, that is, $h$ and the value of
$\delta_{T}$ implicit in the choice of $d^{f}_{T}$.

An optimal value of $h$ can be achieved by selecting the order of the
autoregression using AIC, or an equivalent criterion. Denote the said estimate
by $\widehat{h}_{AIC}$.
Then $\widehat{h}_{AIC}=\mathrm{argmin}_{h=0,1,\ldots,H_{T}}\ln(\hat{\sigma
}_{h}^{2})+2h/T$, where $\hat{\sigma}_{h}^{2}$ denotes the residual mean
square obtained from a fitted $AR(h)$ model. Let $\bar{h}_{T}=\mathrm{argmin}%
_{h=0,1,\ldots,H_{T}}L_{T}(h)$ where $L_{T}(h)=(\sigma_{h}^{2}-\sigma
^{2})+h\sigma^{2}/T$ and $\sigma^{2}$ and $\sigma_{h}^{2}$ are as defined in
Assumption (\ref{Ass2}) and equation \eqref{Varh} respectively. The function
$L_{T}(h)$ was introduced by \cite{shibata:1980} as a figure of merit and the
$AR(\widehat{h}_{AIC})$ model is asymptotically efficient in the sense that
$L_{T}(\widehat{h}_{AIC})=L_{T}(\bar{h}_{T})\{1+o(1)\}$ $a.s.$ as
$T\rightarrow\infty$ \citep[][Theorem
9]{poskitt:2007}. It follows that $\widehat{h}_{AIC}/\bar{h}_{T}\rightarrow1 $
$a.s.$ as $T\rightarrow\infty$, so as $T$ increases $\widehat{h}_{AIC}$
behaves like a deterministic sequence and yields an autoregressive order
$h\sim K\log T$ $a.s.$

Appropriate selection of the pre-filtering value for $d$ is less clear. From
Theorem \ref{pfsbbias} we can see that we require $d_{T}^{f}$ to be such that
$|d_{T}^{f}-d|\log T=o(1)$ $a.s.$, but no other features of the result nor its
derivation give us a guide as to suitable choices for $d_{T}^{f}$. If
$N^{1/2}(d_{T}^{f}-d)$ were exactly $\mathbb{N}(0,\upsilon)$ then it would
follow from the tail area properties of the normal distribution that
$\lim_{T\rightarrow\infty}P(|d_{T}^{f}-d|>\epsilon N^{-1/2+\delta})\leq
\exp(-\epsilon^{2}N^{2\delta}/2\upsilon)$ for any $\delta$, $0<\delta<0.5$ and
$\epsilon>0$. Since $\exp(-\epsilon N^{\delta}/2\upsilon)<|r|^{N^{\delta}}$
for all $r$ such that $\exp(-\epsilon/2\upsilon)<|r|<1$ we could then conclude
from the Borel-Cantelli lemma that $N^{1/2-\delta}|d_{T}^{f}-d|$ converged to
zero $a.s.$ It would then follow that $|d_{T}^{f}-d|\log T=o(1)$ $a.s.$ as
required by Theorem \ref{pfsbbias} since $\log T/N^{1/2-\delta}\rightarrow0$
for all $N\sim KT^{\nu}$ where $2/3<\nu<4/5$. Approximate Gaussianity
associated with the pre-filtering value $d_{T}^{f}$ being a $\sqrt{N}$--CAN
estimator of $d$ is not sufficient to establish the required result, however,
because departures of $N^{1/2}(d_{T}^{f}-d)$ from zero that are
inconsequential for weak convergence need not be immaterial for
large-deviation probabilities. Nevertheless, the necessary large-deviation
property can be derived on a case by case basis.

\begin{proposition}
\label{lrgdev} Let $d_{T}^{f}$ denote any one of the estimators LPR, LPR-BA,
SPLW or SPLW-BA. Then under the conditions of Theorem \ref{pfsbbias}
$|d_{T}^{f}-d|\log T\rightarrow0$ as $T\rightarrow\infty$ with probability
one.
\end{proposition}

Proposition \ref{lrgdev} indicates that each of the estimators to be
considered here can serve as a legitimate pre-filtering value, and in the
simulation experiments we choose to set the (initial) pre-filtering value
equal to the actual estimator to be bias-adjusted. Whilst the latter is
perhaps not strictly necessary, it is an obvious choice to make and a choice
that is also consistent with the details of the proof provided in the paper
for the convergence of the bootstrap bias in (\ref{btbar}) to the actual
finite sample bias in (\ref{bt}).\footnote{Specifically, the constant $\beta$
that appears in the expressions (\ref{biasd}) and (\ref{biasbsd}) for $b_{T}$
and $b_{T}^{\ast}$ respectively is common to both expressions only if
$d_{T}^{f}$ is equivalent to the estimator being bias-adjusted. The presence
of this common factor enables the result in Theorem 3.2 to be produced via
convergence arguments concerning the quantities $\left\vert \overline{c}%
_{0}-c_{0}\right\vert $ and $\left\vert \overline{c}_{1}-c_{1}\right\vert .$}
Furthermore, in the context of the bootstrap algorithm, any bootstrap
bias-adjusted version of an initial estimator can serve as a valid
pre-filtering value in a subsequent application of the algorithm. This
observation, in turn, prompts the following adaptation of the PFSB(BA)
algorithm, in which successive bias-adjusted estimators play the role of the
preliminary pre-filtering value within an iterative scheme.

\section{A Recursive Bias Correction Procedure\label{recursive}}

Although the bias of the bias-adjusted estimator $\widetilde{d}_{T}$ in
\eqref{bias_adjusted_est} will be smaller than that of $\widehat{d}_{T}$, any
bias remaining in $E[\widetilde{d}_{T}]-d$ may still be large because the bias
in any preliminary value $d^{f}$ can be severe in finite samples, and
$\widehat{b}_{T,B}^{\ast_{f}}$ will, as a consequence, be a biased estimate of
its true counterpart $b_{T}$. To obtain a more accurate estimate of $d$ we
propose a further refinement to the proposed correction of $\widehat{d}_{T}$
through a recursive algorithm: \

\begin{description}
\item[BA1$^{\prime}.$] Initialization: Set $k=0$ and assign desirable
tolerance levels $\tau_{1}=\tau_{1}^{(0)}$ and $\tau_{2}=\tau_{2}^{(0)}$. For
the chosen estimator $\widehat{d}_{T}$, set $\widetilde{d}_{T}^{(0)}%
=\widehat{d}_{T}$ (i.e. set $d^{f}=$ $\widehat{d}_{T}$). Now go to Step
BA2$^{\prime}$.

\item[BA2$^{\prime}.$] Recursive Calculation: For the $k$th iteration set the
preliminary value of $d$, namely $d^{f},$ to $\widetilde{d}_{T}^{(k)}$ and
repeat Steps BA2 and BA3 of the PFSB(BA) algorithm with $\widehat{d}_{T}$
therein replaced by $\widetilde{d}_{T}^{(k)}$ to give, in an obvious notation,
$\widetilde{d}_{T}^{(k+1)}=\widetilde{d}_{T}^{(k)}-\widetilde{b}_{T,B}%
^{\ast_{f}(k)}$. Proceed to Step BA3$^{\prime}$.

\item[BA3$^{\prime}.$] Stopping Rule: If $|\widetilde{d}_{T}^{(k+1)}%
-\widetilde{d}_{T}^{(k)}|>\tau_{1}$ and $|\widetilde{d}_{T}^{(0)}%
-\widetilde{d}_{T}^{(k)}-\widetilde{b}_{T,B}^{\ast_{f}(k)}|>\tau_{2}$ set
$k=k+1$, update the tolerance levels $\tau_{1}=\tau_{1}^{(k)}$ and $\tau
_{2}=\tau_{2}^{(k)}$, and repeat Step BA2$^{\prime}$. Otherwise set
$\widetilde{d}_{T}=\widetilde{d}_{T}^{(k)}$ and stop.
\end{description}

The rationale behind the recursions is as follows: since the estimator
$d^{f}=\widehat{d}_{T}$ tends to be severely biased, $\widehat{b}_{T,B}%
^{\ast_{f}}$ will on average be a biased estimate of $b_{T}$, and the
bias-adjusted estimate $\widetilde{d}_{T}$ will therefore still contain some
bias. Replacing the initial values $\widehat{d}_{T}=\widetilde{d}_{T}^{(0)}$
and $\widehat{b}_{T,B}^{\ast_{f}}=\widetilde{b}_{T,B}^{\ast_{f}(0)}$ by
$\widetilde{d}_{T}^{(1)}$ and $\widetilde{b}_{T,B}^{\ast_{f}(1)}$, and (for
general $k$) $\widetilde{d}_{T}^{(k-1)}$ and $\widetilde{b}_{T,B}^{\ast
_{f}(k-1)}$ by $\widetilde{d}_{T}^{(k)}$ and $\widetilde{b}_{T,B}^{\ast
_{f}(k)}$, and so on, produces more accurate estimates and bias assessments.
Being based upon more accurate estimators, the updated estimate $\widetilde
{d}_{T}^{(k)}$ would be expected to be closer to the true value of $d$. The
procedure is iterated until no meaningful gain in accuracy is achieved.

To determine if any meaningful gain in accuracy will be achieved by adding a
further iteration, two criteria are used. The first, $|\widetilde{d}%
_{T}^{(k+1)}-\widetilde{d}_{T}^{(k)}|>\tau_{1}^{(k)}$, is based on Cauchy's
convergence criterion. Given the stochastic nature of the bias correction
mechanism we can think of this as a statistical decision rule in which
$\tau_{1}^{(k)}$ governs the probability of moving from the $k$th to the
$(k+1)$th iteration. Now
\begin{align*}
\widetilde{d}_{T}^{(k+1)}-\widetilde{d}_{T}^{(k)}  & =-\widetilde{b}%
_{T,B}^{\ast_{f}(k)}\\
& =\widetilde{d}_{T}^{(k)}-\frac{1}{B}%
{\textstyle\sum\limits_{b=1}^{B}}
\tilde{d}_{T,b}^{\ast_{f}(k)}\\
& =-\frac{1}{B}%
{\textstyle\sum\limits_{b=1}^{B}}
\left(  \tilde{d}_{T,b}^{\ast_{f}(k)}-\widetilde{d}_{T}^{(k)}\right)
\end{align*}
and since $\widehat{d}_{T}$ is a $\sqrt{N}$--CAN estimator, given the data and
the current and previous bootstrap iterations, $N^{\half}(\tilde{d}%
_{T,b}^{\ast_{f}(k)}-\widetilde{d}_{T}^{(k)})\overset{\mathcal{D}}%
{\rightarrow}N(0,\upsilon^{2})$, where $\tilde{d}_{T,b}^{\ast_{f}(k)}%
$\ denotes the estimator produced from a bootstrap draw based on the PFSB(BA)
algorithm, with $\widetilde{d}_{T}^{(k)}$ used as the pre-filtering value. The
conditional (asymptotic) variance of $B^{-1}%
{\textstyle\sum\limits_{b=1}^{B}}
\left(  \tilde{d}_{T,b}^{\ast_{f}(k)}-\widetilde{d}_{T}^{(k)}\right)  $ is
therefore $\upsilon^{2}/NB$, and using the rule that the overall variance
equals the variance of the conditional mean (in this case $Var[\widetilde
{d}_{T}^{(k)}]$) plus the expectation of the conditional variance (in this
case the constant $\upsilon^{2}/NB$) we can infer that the (asymptotic)
variance of the difference between successive bias-adjusted estimators is
given by
\[
Var[\widetilde{d}_{T}^{(k+1)}-\widetilde{d}_{T}^{(k)}]=Var[\widetilde{d}%
_{T}^{(k)}]+\frac{\upsilon^{2}}{NB}\,.
\]
Furthermore, from the recurrence formula
\begin{align*}
\widetilde{d}_{T}^{(k)}  & =\widetilde{d}_{T}^{(k-1)}-\widetilde{b}%
_{T,B}^{\ast_{f}(k-1)}\\
& =\widetilde{d}_{T}^{(k-1)}-\frac{1}{B}%
{\textstyle\sum\limits_{b=1}^{B}}
\left(  \tilde{d}_{T,b}^{\ast_{f}(k-1)}-\widetilde{d}_{T}^{(k-1)}\right)
\end{align*}
it follows by a similar logic that
\[
Var[\widetilde{d}_{T}^{(k)}]=2\cdot Var[\widetilde{d}_{T}^{(k-1)}%
]+\frac{\upsilon^{2}}{NB},
\]
where $Var[\widetilde{d}_{T}^{(1)}]=2\cdot Var[\widetilde{d}_{T}%
^{(0)}]+\upsilon^{2}/NB=(2B+1)\upsilon^{2}/NB$. Moreover, at each iteration
the bias-adjusted estimate is constructed as a linear combination of
asymptotically normal random variables and is itself therefore asymptotically
normal. This indicates that $\tau_{1}^{(k)}$ can be evaluated from percentile
points of the normal approximation.

Similarly, the second convergence criterion, $|\widetilde{d}_{T}%
^{(0)}-\widetilde{d}_{T}^{(k)}-\widetilde{b}_{T,B}^{\ast_{f}(k)}|>\tau
_{2}^{(k)}$, is perhaps best thought of as the decision rule that examines the
difference between the current accumulated bias correction, $\widetilde{d}%
_{T}^{(0)}-\widetilde{d}_{T}^{(k)}$, and the current bootstrap estimate of the
bias, $\widetilde{b}_{T,B}^{\ast_{f}(k)}$. From the expression
\[
\widetilde{d}_{T}^{(0)}-\widetilde{d}_{T}^{(k)}-\widetilde{b}_{T,B}^{\ast
_{f}(k)}=\widetilde{d}_{T}^{(0)}-\left(  \frac{1}{B}%
{\textstyle\sum\limits_{b=1}^{B}}
\tilde{d}_{T,b}^{\ast_{f}(k)}\right)  ,
\]
it follows that the (asymptotic) variance,
\[
Var[\widetilde{d}_{T}^{(0)}-\widetilde{d}_{T}^{(k)}-\widetilde{b}_{T,B}%
^{\ast_{f}(k)}]=\frac{\upsilon^{2}}{N}\left( 1+2^{k-1}\left[ 1+\frac{1}%
{B}\right] \right) \,,
\]
and the tolerance level $\tau_{2}^{(k)}$ can once again be set using
percentile points from the asymptotic normal approximation.

The interpretation of the convergence criteria as statistical decision rules
in which the tolerance levels govern the probability of going from the current
to the next iteration suggests that $\tau_{1}^{(k)}$ and $\tau_{2}^{(k)}$ be
set by reference to conventional critical values used in statistical
hypothesis tests. When $k$ is very small we might conjecture that
$\widetilde{d}_{T}^{(k)}$ still contains some bias and we may wish to iterate
further unless there is strong evidence that so doing will produce very little
change. On the other hand, when $k$ is large the initial estimate
$\widetilde{d}_{T}^{(0)}$ has already undergone several adjustments to produce
$\widetilde{d}_{T}^{(k)}$ and we may prefer to terminate iteration unless
there is strong evidence that further iteration will produce additional,
substantial correction. We can therefore calibrate $\tau_{1}^{(k)}$ and
$\tau_{2}^{(k)}$ using quantile points of the normal distribution
$z_{(1-p_{k}/2)}$ (where $G(z_{(1-p)})=1-p$) and $p_{k}$, the probability of
going from the $k$th to the $(k+1)$th iteration, is assigned to be large when
$k$ is small and vice versa. In the experiments that follow we set
$p_{0}=0.95$, $p_{1}=0.9$, and $p_{k}=(0.1)2^{(1-k)}$ for $k=2,3,\ldots
$\thinspace for uncorrected LPR and SPLW; and $p_{0}=0.9$, $p_{k}=(0.1)2^{-k}$
for $k=1,2,3,\ldots$\thinspace for LPR-BA and SPLW-BA with $P=1$. We comment
further on the stochastic stopping rules when discussing our experimental
results below.

\section{Simulation Exercise\label{sim}}

\subsection{Simulation Design\label{design}}

In this section we illustrate the performance of the bootstrap bias-corrected
estimators via a small simulation experiment. Following
\cite{andrews:guggenberger:2003} we simulate data from a Gaussian
$ARFIMA(1,d,0)$ process,
\begin{equation}
(1-L)^{d}\Phi(z)y(t)=\varepsilon(t)\,,\ 0<d<0.5\,,\label{arfima}%
\end{equation}
where $\Phi(z)=1-\phi z$ is the operator for a stationary AR(1) component and
$\varepsilon(t)$ is zero-mean Gaussian white noise. The choice of this model
is motivated, in part, by earlier work that highlights the distinct finite
sample bias of the LPR estimator of $d$ in this setting, when the value of
$\phi$ is positive and large \citep[See][]{agiakloglou:newbold:wohar:1993}.
Indeed, \cite{andrews:guggenberger:2003} document substantial remaining bias
in the bias-corrected version of the LPR estimator in the presence of a large
autoregressive parameter. That is, the impetus for applying bootstrap-based
bias corrections to the various estimators is particularly strong in this setting.

The process in (\ref{arfima}) is simulated $R=1000$ times for
$d=0.0,0.2,0.3,0.4$; $\phi=0.3,0.6,0.9,$ and sample sizes $T=100,200,500$ via
Levinson recursion applied to the autocovariance function (ACF) of the desired
$ARFIMA(p,d,q)$ process and the generated pseudo-random $\varepsilon(t)$
\citep[see, for instance,][\S5.2]{brockwell:davis:1991}. The ARFIMA ACF for
given $T$, $\phi$, $\theta$, and $d$ is calculated using Sowell's
\citeyearpar{sowell:1992} algorithm as modified by \cite{doornik:ooms:2003}.

The estimators that we bias correct via the iterative PFSB(BA) algorithm are:
LPR, LPR-BA, SPLW and SPLW-BA, implemented with a bandwidth $N=T^{0.7}$ and
$B=1000.$ Values of $P=1,2,$ are used for defining the two (analytically)
bias-adjusted methods. For the log-periodogram regression estimators,
\begin{equation}
N^{\half}(\widehat{d}_{T}-d)\overset{\mathcal{D}}{\rightarrow}N\left(
0,\frac{\pi^{2}}{24}\upsilon_{P}^{2}\right)  \,,\label{lpr}%
\end{equation}
where $\upsilon_{P}^{2}$ gives the variance inflation factor of the estimator.
The inflation factor results from the modeling of $\log|\kappa(e^{-\imath
\lambda})|^{2}$ by a polynomial of degree $2P$ that underlies the bias
correction. For the local polynomial Whittle estimators,
\begin{equation}
N^{\half}(\widehat{d}_{T}-d)\overset{\mathcal{D}}{\rightarrow}N\left(
0,\frac{1}{4}\upsilon_{P}^{2}\right)  \,.\label{splw}%
\end{equation}
For both the LPR and the SPLW estimators the variance inflation factors are
$\upsilon_{0}^{2}=1,\upsilon_{1}^{2}=2.25$ and $\upsilon_{2}^{2}=3.52$
where $\upsilon_{0}^{2}=1$ yields the baseline variance for the uncorrected
estimator, see \cite{andrews:guggenberger:2003} and \cite{andrew:sun:2004}.
The estimators are known to be rate optimal when $N\sim KT^{4/5}$ in the
uncorrected case \citep{giraitis:robinson:samarov:1997} and $N\sim
KT^{(4+4P)/(5+4P)}$ in the corrected case
\citep{andrews:guggenberger:2003,andrew:sun:2004}, but in practice optimal
bandwidths seem not to be used much, the values $N=T^{2/5},T^{\half},T^{3/5}$
and $T^{7/10}$ being popular choices. The order ($h$) of the autoregressive
approximation underlying the sieve component of the bootstrap algorithm is
chosen via $AIC$, and Burg's algorithm is used to estimate the autoregressive
parameters.

Based on the $R$ replications, for each estimator of $d,$ we report the bias
and mean square error (MSE). For comparative purposes, we also document the
sampling performance of the unadjusted estimators (LPR, SPLW) and the
estimators that are analytically adjusted (only) (LPR-BA and SPLW-BA;
$P=1,2$). That is, we are interested, in particular, in documenting: 1) any
improvement that can be had by using the bootstrap method \textit{rather} than
an analytical method to bias correct a given estimator; and 2) any
\textit{additional} improvement associated with bias correcting (via the
bootstrap) an estimator that has already been bias corrected via analytical means.

For each estimator considered (i.e. each of the two base estimators, LPR and
SPLW, and all of the analytically and bootstrap bias-corrected versions
thereof), we also document the empirical coverage (over the Monte Carlo
replications) of the nominal 95\% confidence intervals, plus the average
length of the given intervals. The 95\% confidence intervals (CIs) are
constructed from each of the $R$ bootstrap distributions (each, in turn, based
on $B$ bootstrap draws) as: $\left\{  \widetilde{d}_{T}(L),\widetilde{d}%
_{T}(U)\right\}  ,$ where $\widetilde{d}_{T}(L)$ ($\widetilde{d}_{T}(U)$)
denotes the lower (upper) bound of a highest density interval, in which the
narrowest interval with 95\% coverage for the bootstrap distribution is
selected. For any given estimator $\widetilde{d}_{T},$ empirical coverage for
each interval type is calculated as the proportion of times (in $R$
replications) that each interval covers the true value of $d.$ The average
length of each interval (across the $R$ replications) is also recorded. These
coverage and length statistics for the bootstrap-based estimators are compared
with the empirical coverage and (constant) length of 95\% intervals
constructed for the unadjusted (or analytically-adjusted) LPR and SPLW
estimators, as based on the appropriate asymptotic distributions, in
(\ref{lpr}) and (\ref{splw}) respectively. Note that the value of $B$ used
here implies, from the Dvoretsky--Kiefer--Wolfowitz inequality, that
$\textmd{P}(\sup_{x}|\overline{F}_{\widetilde{d}_{T},B}^{\ast}%
(x)-F_{\widetilde{d}_{T}}^{\ast}(x)|>\delta)<2\exp(-\delta^{2}(1000))$, where
$\overline{F}_{\widetilde{d}_{T},B}^{\ast}(x)$ is the empirical (bootstrap)
distribution of $\widetilde{d}_{T}$, based on $B$ bootstrap draws, and
$F_{\widetilde{d}_{T}}^{\ast}(x)$ is the distribution of $\widetilde{d}_{T}$
under the probability law induced by the bootstrap.

We record results for the bootstrap-based estimators produced through formal
application of the stopping rules described above. To the two stochastic
stopping criteria we add a deterministic criterion, whereby the iterative
scheme ceases if $\widetilde{d}_{T}^{(k+1)}<-1$ or $\geq1.5$ and the estimator
$\widetilde{d}_{T}^{(k)}$ retained as the final choice. We also record results
for the estimators based on only one and two iterations of the iterative
method ($k=1,2$ in Steps BA2$^{\prime}$ and BA3$^{\prime}$). In the following
section we discuss all numerical results associated with the LPR estimator,
and Section \ref{splw_results} all results for the SPLW estimator, with the
relevant tables included in Appendix B. Note that most results for
$T=200$\ and $d=0.3$\ are omitted for brevity. The coverage and length results
for the three different values of $\phi$ are reported after averaging over all
four values of $d$, including $d=0.3.$

\subsection{Simulation Results:\ LPR}

Tables \ref{table_lpr_bias_100} and \ref{table_lpr_bias_500} record (for
$T=100$ and $500$ respectively) the bias and MSE results for all estimators
based on the LPR method. All results pertaining to the use of the
bootstrapping to bias adjust are indicated by the subscript `$sb$' appearing
on the relevant acronym for the estimator (LPR or LPR-BA), both in the
subsequent text and the tables. In all tables the most favorable result for
each parameter setting is highlighted in bold. The columns headed `SSR' in the
tables report the results based on the stochastic stopping rules discussed in
Section \ref{recursive} and modified (deterministically) as described at the
end of Section \ref{design}.

The key message to be gleaned from the numerical results presented in Tables
\ref{table_lpr_bias_100} and \ref{table_lpr_bias_500} is that the bootstrap
technique \textit{does }reduce bias, but with the most substantial gains to be
had by using the bootstrap algorithm to bias-adjust an estimator that has
already been bias adjusted analytically. For example, for $T=100\,$, and for
$\phi=0.3,0.6$, in all but one of the six cases, the smallest bias is produced
by bias adjusting (via the bootstrap) the LPR-BA($P=2$) estimator once, with
no subsequent iteration: LPR-BA$_{sb}$($P=2,k=0)$. For $T=500$, this estimator
is the least biased estimator for \textit{all} three values of $d$ and for
$\phi=0.3,0.6.$ Importantly, for these two values of $\phi$ (and for both
sample sizes) if one compares the MSE of LPR-BA$_{sb}$($P=2,k=0)$ with that of
LPR-BA($P=3)$, the reduction in bias produced by the bootstrap technique is
\textit{not}\textbf{\ }obtained at the expense of MSE, with the two estimators
having very similar MSE's, and one not systematically dominating the other in
terms of this performance measure. For $\phi=0.9$, \textit{all }versions of
the LPR estimator, including the bootstrap bias-adjusted versions, are very
biased. That said, for $T=500$, the estimator with the \textit{smallest} bias
is the raw LPR ($P=0$) estimator bootstrap bias-adjusted three times:
LPR$_{sb}(k=2$).

A detailed examination of the simulation outcomes indicates that the
stochastic stopping rules usually terminate the iterative procedure after
zero, one or two iterations, with evidence for this provided by the nature of
the bias and MSE results recorded in Tables \ref{table_lpr_bias_100} and
\ref{table_lpr_bias_500}. Looking first at the results in the middle panel of
Table \ref{table_lpr_bias_100}, we see that bias in the SSR column falls
between the bias recorded for $k=1$ and $k=2$ respectively. The same
observation can be made for the MSE. With the exception of the MSE results for
$\phi=0.9$, the same conclusion can be drawn for the results recorded in the
middle panel of Table \ref{table_lpr_bias_500} for $T=500.$ For the cases
where an analytical bias adjustment precedes the iterative bootstrap procedure
(as recorded in the third panel of Tables \ref{table_lpr_bias_100} and
\ref{table_lpr_bias_500}) we find that the bias and MSE recorded in the SSR
column almost always fall between the comparable results for $k=0$ and $k=1.$
Hence, we can conclude that although a stochastic stopping rule tailors the
number of iterations to the realization at hand, its use does not appear to
\textit{guarantee} an improvement in overall performance compared to using a
fixed number of iterations.

Table \ref{table_lpr_HPD} summarizes the empirical coverage performance of
highest probability density (HPD) confidence intervals for the alternative
estimators, for both sample sizes and based on a nominal coverage of 95\%. The
second panel of this table records the average length (across simulations) of
the 95\% HPD intervals for all cases. Coverage (and length) results for the
nominal level of 90\% are qualitatively similar, and hence are not reported.

In terms of coverage, a combination of analytical and bootstrap-based bias
adjustments once again yields the best results overall, with either
LPR-BA$_{sb}$($P=1,k=1$) (i.e. LPR-BA$(P=1)$ bootstrap bias-adjusted twice) or
LPR-BA$_{sb}$($P=2,k=0$) (i.e. LPR-BA$(P=2)$ bootstrap bias-adjusted once)
having the best empirical coverage -- and very \textit{accurate} empirical
coverage -- in all four cases recorded in Table \ref{table_lpr_HPD} for
$\phi=0.3,0.6.$ Once again, all coverage results for $\phi=0.9$ are poor,
although, for what it is worth, for $T=500$, the bootstrapped bias-adjusted
LPR-BA ($P=2$) produces the most accurate coverage interval (at 32\%!).

In terms of the length of the 95\% intervals, there are two key points to
note. Firstly, it is the asymptotic intervals that are the most narrow, but
this precision is at the expense of very inaccurate coverage. Secondly, the
coverage accuracy yielded by the bootstrap is \textit{not }at the expense of
precision. That is, any bootstrap-based bias correction that improves coverage
produces a negligible change in the length of the interval. This result
provides an interesting contrast with the corresponding results for analytical
bias-adjustment; i.e. any such analytical adjustment that improves coverage
does so at the expense of a decrease in precision, with the 95\% intervals
widening as the value of $P$ increases.

This raises the question of how the sieve bootstrap is able to bias correct
the LPR or a LPR-BA estimator without incurring any loss of precision. The
motivation underlying log-periodogram regression is that
\begin{equation}
\frac{I_{T}(\lambda)2\pi|1-e^{-\imath\lambda}|^{2d}}{\sigma^{2}|\kappa
(e^{\imath\lambda})|^{2}}\overset{\mathcal{D}}{\rightarrow}%
Exp(1),\label{perio}%
\end{equation}
and using the approximation $|1-e^{-\imath\lambda}|^{2d}=|\lambda
|^{2d}(1+o(1))$ as $\lambda\rightarrow0$ we have the linear regression model
\begin{equation}
\log(I_{T}(\lambda_{j}))=\alpha_{0}-2d\log(\lambda_{j})+\eta_{j},\label{lpr1}%
\end{equation}
where $E[\eta_{j}]=0$ and the intercept $\alpha_{0}$ is presumed to capture
the effects of the adjustments
\begin{align}
a_{j}  & =\log|\kappa(1)|^{2}+\log\left(  \frac{|\kappa(e^{\imath\lambda_{j}%
})|^{2}}{|\kappa(1)|^{2}}\right)  -d\log\left(  \frac{|1-e^{-\imath\lambda
_{j}}|^{2}}{\lambda_{j}^{2}}\right)  -C\label{lpr2}\\
& =\log|\kappa(1)|^{2}-C+O(N^{2}/T^{2})\quad\mbox{for all}\quad1\leq j\leq
N~,\label{lpr2b}%
\end{align}
where $C=0.577216$ (Euler's constant).\footnote{The expression in
\eqref{lpr2b} follows as a consequence of the fact that $\log(|\kappa
(e^{\imath\lambda})|^{2}/|\kappa(1)|^{2})=\log(1+(c_{1}/c_{0})|\lambda
|^{2}+o(|\lambda|^{3}))$ and $\log(|1-e^{-\imath\lambda}|^{2}/\lambda
^{2})=\log(1-(1/12)|\lambda|^{2}+o(|\lambda|^{3}))$ as $\lambda\rightarrow0$.}
The presumption that $\alpha_{0}$ absorbs the effects of the adjustment term
assumes $a_{j}$ approaches $\log|\kappa(1)|^{2}-C$ sufficiently quickly that
the deviations $a_{j}-\log|\kappa(1)|^{2}+C$ can be ignored.

The analytical correction replaces the simple regression in \eqref{lpr1} by
the multiple regression
\begin{equation}
\label{lpr3}\log(I_{T}(\lambda_{j}))=\sum_{p=0}^{P}\alpha_{p}\lambda_{j}%
^{2p}-2d\log(\lambda_{j})+\eta_{j}\,,
\end{equation}
the rationale being that the term $\sum_{p=0}^{P}\alpha_{p}\lambda_{j}^{2p} $
provides a better approximation to the Maclaurin series expansion of the right
hand side of \eqref{lpr2} than supposing $a_{j}$ is constant in a
neighbourhood of zero. The introduction of $\lambda_{j}^{2p}$, $p=1,\ldots,P$,
in \eqref{lpr3} reduces the bias in the estimate of $d$, but it is also the
presence of these additional regressors that causes the variance inflation
seen in \eqref{lpr}.

The PFSB, on the other hand, takes the specification of the regression in
\eqref{lpr1} or \eqref{lpr3} as given and adjusts the estimator by mimicking
the sampling behaviour of the regressand.
Recall that $I_{T}(\lambda)=(2\pi)^{-1}\sum_{r=1-T}^{T-1}\widehat{\gamma
}(r)e^{\imath\lambda r }$. \cite{hosking:1996} shows that when $d$ is large
the $\widehat{\gamma}(r)$ have substantial negative bias relative to the true
autocovariances, even for moderate to large samples.
The PFSB reduces the memory in the \textquotedblleft data\textquotedblright%
\ to which the sieve bootstrap is applied, via the pre--filtering procedure,
so as to give a near optimal convergence rate when implicitly assessing the
corresponding bias in $\log(I_{T}(\lambda))$. Whether it is applied to
\eqref{lpr1} or \eqref{lpr3}, the PFSB is thereby able to attack the problem
of bias in the estimation of $d$ without compromising the pivotal nature of
the ratio in \eqref{perio}, the basic result that underlies the
log-periodogram regressions and determines the estimators' variance.

\subsection{Simulation Results: SPLW\label{splw_results}}

Tables \ref{table_lpw_bias_100} and \ref{table_lpw_bias_500} record (for
$T=100$ and $500$ respectively) the bias and MSE results for all estimators
based on the SPLW method (with the subscript `$sb$' used as descibed above),
whilst Table \ref{table_lpw_HPD} records the 95\% interval coverage and length
statistics, for all cases. Once again, the most favorable result for each
parameter setting is highlighted in bold in all tables. As with the LPR-based
estimators, the bootstrap-based bias adjustment yields the largest bias
reductions, but only when applied to an SPLW estimator that has already been
analytically bias adjusted. In contrast with the LPR-based results, these bias
gains are evident only for the larger of the two sample sizes ($T=500$), with
there being no gain (over full analytical adjustments) in the $T=100$ case.
The bias gains (for the $T=500$ case) are for $\phi=0.3,0.6$ only, with the
least biased estimator for $\phi=0.9$ being the SPLW-BA ($P=3 $) estimator.
The biases of all SPLW-based estimators are similar to the biases of the
comparable LPR-based estimators, and as with the LPR-based estimators, the
reduction in bias produced by the bootstrap technique (in certain cases) is
not\textbf{\ }obtained at the expense of MSE. Once again, although the use of
a stochastic stopping rule is appealing, as was the case for the LPR results
it does not guarantee an improvement in performance over using a fixed number
of iterations.

The coverage results for the SPLW-based estimators are qualitatively identical
to those for the LPR case; in particular, the bootstrap bias adjustment of an
already analytically adjusted estimator yields the best coverage for
$\phi=0.3,0.6$, for both sample sizes - and very accurate coverage at that.
Although the bootstrapped bias-adjustment of LPR-BA ($P=2$) produces the most
accurate coverage for the $\phi=0.9$ case (for both sample sizes), the
coverage results are poor for all estimators in this part of the parameter
space. As for the LPR case, the bootstrap-based bias adjustment is not
accompanied by an increase in interval length, in contrast with the analytical
bias adjustment. As a consequence, the bootstrap method can be used to yield
coverage that is close to the nominal level without sacrificing inferential precision.

\section{Conclusion}

This paper has developed a bootstrap method for bias correcting
semi-parametric estimators of the long memory parameter in fractionally
integrated processes. The method involves applying the sieve bootstrap to data
pre-filtered by a preliminary semi-parametric estimate of the long memory
parameter. In addition to providing theoretical (asymptotic) justification for
using the bootstrap techniques, we document the results of simulation
experiments, in which the finite sample performance of the (bias-adjusted)
estimators is compared with that of both unadjusted estimators and estimators
adjusted via analytical means. The numerical results are very encouraging, and
suggest that the bootstrap bias correction \textit{can }yield more accurate
inferences about long memory dynamics in the types of samples that are
encountered in practice.

\appendix

\makeatletter\def\@seccntformat#1{\csname Pref@#1\endcsname\csname
the#1\endcsname: \ }
\def\Pref@section{Appendix~}
\makeatother

\section{Proofs}

\label{app1}

\paragraph{Proof of Theorem \ref{consistentYW}:}

For the \LS \ and \YW \ estimators see \citet[][Theorem 5 and Corollary
1]{poskitt:2007} and the associated discussion. For the Burg estimator the
result then follows from \citet[][Theorem 1]{poskitt:1994}.$\hfill\qed$

\paragraph{Proof of Theorem \ref{Biasadj}:}

Subtracting \eqref{Chib} from \eqref{Chib*} and using the triangular
inequality we find that $|\textmd{P}^{\ast}\{N^{\half}(\widehat{d}_{T}%
^{\ast_{f}}-E^{\ast}[\widehat{d}_{T}^{\ast_{f}}])<x\}-\textmd{P}%
\{N^{\half}(\widehat{d}_{T}-E[\widehat{d}_{T}])<x\}|$ is less than or equal
to
\[
|G((x+N^{\half}b_{T})/\upsilon)-G((x+N^{\half}b_{T}^{\ast})/\upsilon
)|+o(N^{-\half})\,.
\]
But
\[
\sup_{x}|G((x+N^{\half}b_{T})/\upsilon)-G((x+N^{\half}b_{T}^{\ast}%
)/\upsilon)|\leq\frac{N^{\half}}{\upsilon\sqrt{2\pi}}|b_{T}-b_{T}^{\ast}|
\]
by the first mean value theorem for integrals \citep[][Theorem
7.30]{apostol:1960} and the theorem follows.$\hfill\qed$

\paragraph{Proof of Lemma \ref{kk}:}

Trivial addition and subtraction yields
\begin{align}
\label{kkbar}|\bar{\kappa}_{h}(e^{\imath\lambda})|^{2}-|\kappa(e^{\imath
\lambda})|^{2}=(|\bar{\kappa}_{h}(e^{\imath\lambda})|^{2}- & |\kappa^{f}%
_{h}(e^{\imath\lambda})|^{2})+(|\kappa^{f}_{h}(e^{\imath\lambda})|^{2}%
-|\kappa^{f}(e^{\imath\lambda})|^{2})\nonumber\\
& +(|\kappa^{f}(e^{\imath\lambda})|^{2}-|\kappa(e^{\imath\lambda})|^{2})\,.
\end{align}

Consider the first term in \eqref{kkbar}, $|\bar{\kappa}_{h}(e^{\imath\lambda
})|^{2}-|\kappa_{h}^{f}(e^{\imath\lambda})|^{2}$. By definition
\[
\bar{\kappa}_{h}(z)-\kappa_{h}^{f}(z)=\frac{\phi_{h}^{f}(z)-\bar{\phi}_{h}%
(z)}{\bar{\phi}_{h}(z)\phi_{h}^{f}(z)}\,,
\]
and since $\bar{\phi}_{h}(z)\neq0$ and $\phi_{h}^{f}(z)\neq0$, $|z|\leq1$,
there exists an $\epsilon>0$ such that
\begin{align*}
|\bar{\kappa}_{h}(z)-\kappa_{h}^{f}(z)|  & \leq\epsilon^{-2}|\phi_{h}%
^{f}(z)-\bar{\phi}_{h}(z)|\\
& \leq\epsilon^{-2}\sum_{j=0}^{h}|\phi_{h}^{f}(j)-\bar{\phi}_{h}%
(j)|\quad\text{for all}\quad|z|\leq1\,.
\end{align*}
But
\begin{align*}
\sum_{j=0}^{h}|\phi_{h}^{f}(j)-\bar{\phi}_{h}(j)|  & \leq\left(  h\sum
_{j=0}^{h}|\phi_{h}^{f}(j)-\bar{\phi}_{h}(j)|^{2}\right)  ^{\half}\\
& =O\left(  h\left(  \frac{\log T}{T}\right)  ^{\half(1-2\max\{0,d-d^{f}%
\})}\right) \\
& =O\left(  h\left(  \frac{\log T}{T}\right)  ^{\half-\delta_{T}}\right)
\quad a.s.
\end{align*}
by Theorem \ref{consistentYW} and the fact that $|d^{f}-d|<\delta_{T}$ by
assumption. It follows that $|\bar{\kappa}_{h}(e^{\imath\lambda})-\kappa
_{h}^{f}(e^{\imath\lambda})|=O(h(\log T/T)^{\half-\delta_{T}})$ $a.s.$
uniformly in $\lambda$, and hence that $\left\vert |\bar{\kappa}_{h}%
(e^{\imath\lambda})|^{2}-|\kappa_{h}^{f}(e^{\imath\lambda})|^{2}\right\vert
=O(h(\log T/T)^{\half-\delta_{T}})$ $a.s.$ uniformly in $\lambda$. We can
therefore interchange limit operations \citep[][Theorem
13.3]{apostol:1960} to give
\[
\lim_{T\rightarrow\infty}\lim_{\lambda\rightarrow0}\left\vert |\bar{\kappa
}_{h}(e^{\imath\lambda})|^{2}-|\kappa_{h}^{f}(e^{\imath\lambda})|^{2}%
\right\vert =\lim_{\lambda\rightarrow0}\lim_{T\rightarrow\infty}\left\vert
|\bar{\kappa}_{h}(e^{\imath\lambda})|^{2}-|\kappa_{h}^{f}(e^{\imath\lambda
})|^{2}\right\vert \,,
\]
which implies that $\nu_{1,T}=O(h(\log T/T)^{\half-\delta_{T}})$ $a.s.$ for
all $\lambda\in\lbrack2\pi/T,2\pi N/T]$.

For the second term in \eqref{kkbar}, $|\kappa^{f}_{h}(\rho()|^{2}-|\kappa
^{f}(e^{\imath\lambda})|^{2}$, we have
\[
\kappa^{f}_{h}(z)-\kappa^{f}(z) = \frac{1-\kappa^{f}(z)\phi^{f}_{h}(z)}%
{\phi^{f}_{h}(z)}\,,
\]
giving us the bound
\[
|\kappa^{f}_{h}(z)-\kappa^{f}(z)|\leq\epsilon^{-1}|1-\kappa^{f}(z)\phi^{f}%
_{h}(z)|\quad\text{for all}\quad|z|\leq1\,.
\]
Let $\rho_{h}(z)=\sum_{j\geq1}\rho_{h}(j)z^{j}=1-\kappa^{f}(z)\phi^{f}_{h}%
(z)$. Then from Parseval's relation
\[
\sum_{j\geq1}\rho_{h}(j)^{2}=\int_{-\pi}^{\pi}|1-\kappa^{f}(e^{\imath\lambda
})\phi^{f}_{h}(e^{\imath\lambda})|^{2}d\lambda=2\pi\sigma^{-2}(\sigma^{f2}%
_{h}-\sigma^{2})
\]
and from the Levinson--Durbin recursions \citep{levinson:1947,durbin:1960} we
have $\sigma^{f2}_{h}=(1-\phi^{f}_{h} (h)^{2})\sigma_{h-1}^{f2}$. Substituting
sequentially in the recurrence formula $\sigma_{h}^{f2}=\sigma_{h+1}^{f2}%
+\phi^{f}_{h} (h)^{2}\sigma_{h}^{f2}$ leads to the series expansion
$\sigma^{f2}_{h}-\sigma^{2}= \sum_{r=h}^{\infty}\phi^{f}_{r} (r)^{2}\sigma
_{r}^{f2}$, from which we obtain the bound
\[
\sum_{j\geq1}\rho_{h}(j)^{2}\leq2\pi\sigma^{-2}E[w^{f}(t)^{2}]\sum
_{r=h}^{\infty}\phi^{f}_{r} (r)^{2}\,.
\]
But $\phi^{f}_{h} (h)\sim|d-d^{f}|/h$ as $h\rightarrow\infty$
\citep{inoue:2002,inoue:kasahara:2004} and therefore we can infer that
\[
\sum_{j\geq1}\rho_{h}(j)^{2}\leq\text{const.}\frac{|d-d^{f}|^{2}}{h^{2|d|}}
\zeta(2(1-|d|)),
\]
where $\zeta(\cdot)$ denotes the Riemann zeta function. It follows that
$\lim_{h\rightarrow\infty}\rho_{h}(e^{\imath\lambda})=0$ and that
$\lim_{T\rightarrow\infty}|\rho_{h}(e^{\imath\lambda})|^{2}=O(\delta_{T}%
^{2}h^{-2|d|}) $ almost everywhere on $[-\pi,\pi]$. Hence we can conclude that
$\lim_{T\rightarrow\infty}\lim_{\lambda\rightarrow0}\left| |\kappa^{f}%
_{h}(e^{\imath\lambda})|^{2}-| \kappa^{f}(e^{\imath\lambda})|^{2}\right|
=\lim_{\lambda\rightarrow0}\lim_{T\rightarrow\infty}\left| |\kappa^{f}%
_{h}(e^{\imath\lambda})|^{2}-| \kappa^{f}(e^{\imath\lambda})|^{2}\right| $ and
hence that $\nu_{2,T}=O(\delta_{T}^{2}h^{-2|d|}) $.

The third and final term in \eqref{kkbar} is
\begin{equation}
|\kappa^{f}(e^{\imath\lambda})|^{2}-|\kappa(e^{\imath\lambda})|^{2}%
=|\kappa(e^{\imath\lambda})|^{2}(|1-e^{\imath\lambda}|^{2(d^{f}-d)}%
-1)\,.\label{kk3}%
\end{equation}
Substituting $|1-e^{\imath\lambda}|^{2(d^{f}-d)}=\exp\{(d^{f}-d)\log
|1-e^{\imath\lambda}|^{2}\}$ into \eqref{kk3} and using the expansion
$|1-e^{-\imath\lambda}|^{2}=2\sum_{j=1}^{\infty}(-1)^{j-1}|\lambda
|^{2j}/(2j)!$, which implies that $\log|1-e^{\imath\lambda}|^{2}=2\log
|\lambda|+\log(1+o(|\lambda|))$ as $\lambda\rightarrow0$, we can deduce that
\[
\left\vert |\kappa(e^{\imath\lambda})|^{2}(|1-e^{\imath\lambda}|^{2(d^{f}%
-d)}-1)\right\vert \leq\{\sup_{[-\pi,\pi]}|\kappa(e^{\imath\lambda}%
)|^{2}\}|\exp\{2(d^{f}-d)\log|\lambda|+o(|\lambda|)\}-1|
\]
as $\lambda\rightarrow0$. Furthermore, by assumption $|d^{f}-d|\leq\delta_{T}$
where $\delta_{T}\log T\rightarrow0$ as $T\rightarrow\infty$, and since
$|\exp(x)-1|=|x|\cdot|1+\half x+o(|x|)|$ for $x$ in a neighbourhood of the
origin, it follows that
\[
\left\vert |\kappa(e^{\imath\lambda})|^{2}(|1-e^{\imath\lambda}|^{2(d^{f}%
-d)}-1)\right\vert \leq2\{\sup_{[-\pi,\pi]}|\kappa(e^{\imath\lambda}%
)|^{2}\}|d^{f}-d||(\log2\pi N/T)+o(N/T)|
\]
for all $\lambda\in[2\pi/T,2\pi N/T]$ as $T\rightarrow\infty$. We can
therefore infer that \eqref{kk3} is $O(\delta_{T}\log T)$ or smaller,
uniformly in $\lambda$ for all $\lambda\in[ 2\pi/T,2\pi N/T]$. The lemma now
follows. $\hfill\qed$

\paragraph{Proof of Theorem \ref{pfsbbias}:}

It is sufficient to show that $|\bar{c}_{0}-c_{0}|$ and $|\bar{c}_{1}-c_{1}|$
are of order $O\left( T^{2} M_{T}/N^{2}\right) $ or smaller where $M_{T}%
=\max\{h(\frac{\log T}{T})^{\half-\delta_{T}},\delta_{T}h^{-|d|},\delta
_{T}\log T\}$. Evaluating the expression
\begin{equation}
(\bar{c}_{0}-c_{0})+(\bar{c}_{1}-c_{1})|\lambda|^{2}=|\bar{\kappa}%
_{h}(e^{\imath\lambda})|^{2}-|\kappa(e^{\imath\lambda})|^{2}+o(|\lambda
|^{3})\label{pfsbbias2}%
\end{equation}
at $\lambda=2\pi/T$ and $2\pi N/T$, and solving for $\bar{c}_{0}-c_{0}$ and
$\bar{c}_{1}-c_{1}$, it follows a consequence of Lemma \ref{kk} that $|\bar
{c}_{0}-c_{0}|=O\left( M_{T}\right) +o(T^{-3})$ and $|\bar{c}_{1}%
-c_{1}|=O\left( T^{2}M_{T}/N^{2}\right) +o(N/T)$. Extracting the dominant term
gives the desired result. $\hfill\qed$

\paragraph{Proof of Proposition \ref{lrgdev}:}

Let $\widehat{d}_{T}$ denote the LPR estimator. Then $\widehat{d}_{T}$ is the
OLS coefficient of the regressor $-2\log\lambda_{j}$ in the regression of
$\log I_{T}(\lambda_{j})$ on $1$ and $-2\log\lambda_{j} $. Substituting
$a_{j}-2d\log(\lambda_{j})+\eta_{j}$ for $\log I_{T}(\lambda_{j})$ in this
regression leads to the expression
\begin{align}
\label{esterror}\widehat{d}_{T}-d & =-\frac{\sum_{j=1}^{N}(\log\lambda
_{j}-\overline{\log\lambda})(\eta_{j}+a_{j})}{2\sum_{j=1}^{N}(\log\lambda
_{j}-\overline{\log\lambda})^{2}}\nonumber\\
& =-\half\sum_{j=1}^{N}r_{j}(\eta_{j}+a_{j})
\end{align}
for the estimation error where $\eta_{j}$ and $a_{j}$ are defined in
expressions \eqref{lpr1} and \eqref{lpr2}, and $r_{j}=(\log\lambda
_{j}-\overline{\log\lambda})/\sum_{j=1}^{N}(\log\lambda_{j}-\overline
{\log\lambda})^{2}$, $j=1,\ldots,N$. See the discussion associated with
\eqref{lpr1} and \eqref{lpr2} for clarification.

By Theorem 2 of \cite{moulines:soulier:1999} there exists sequences $e_{j}$
and $f_{j}$, $j=1,\ldots,N$, such that $\eta_{j}=e_{j}+f_{j}$, where the
$e_{j}$, $j=1,\ldots,N$, are weakly dependent, centered Gumbel random
variables with variance $\pi^{2}/6$ and covariance $cov\{e_{k},e_{j}%
\}=O(\log^{2}(j)k^{-2|d|}j^{2(|d|-1)})$ for $1\leq k<j\leq N$, and
$|f_{j}|=O(\log(1+j)/j)$ with probability one. Since $\max_{1\leq j\leq
N}|\log\lambda_{j}-\overline{\log\lambda}|=O(\log N)$ and $\sum_{j=1}^{N}%
(\log\lambda_{j}-\overline{\log\lambda})^{2}=O(N)$ it follows that $\sum
_{j=1}^{N}r_{j}f_{j}=O(\log^{3}N/N)$ $a.s.$. Given that $\sum_{j=1}^{N}%
r_{j}=0$, it also follows from \eqref{lpr2b} that $\sum_{j=1}^{N}r_{j}%
a_{j}=O(N^{2}\log N/T^{2})$. We can therefore infer from \eqref{esterror}
that
\[
\widehat{d}_{T}-d=-\half\sum_{j=1}^{N}r_{j}e_{j}+R_{N}%
\]
where $|R_{N}|\log T=O(\nu^{3}\log^{4}T/T^{\nu})+O(\nu\log^{2}T/T^{2(1-\nu
)})=o(1)$ $a.s.$, $2/3<\nu<4/5$.

The desired result now follows because on application of a law of large
numbers for triangular arrays of weakly dependent random variables we find
that for all $\delta>0$
\[
\sum_{j=1}^{N}r_{j}e_{j}=o\left(  (\nu\log T)^{5/2}(\log(\nu\log
T))^{(1+\delta)/2}T^{-\nu/2}\right)  \quad a.s.\quad.
\]
More specifically, let $S_{n}=\sum_{j=1}^{n}r_{j}e_{j}$. Then by Doob's
inequality $E[(\max_{n\leq2^{k}}|S_{n}|)^{2}]\leq4E[|S_{2^{k}}|^{2}]$, and
using the bounds on the covariance of $e_{j}$ we have
\[
E[|S_{n}|^{2}]=\sum_{j=1}^{n}r_{j}^{2}E[e_{j}^{2}]+2\sum_{1\leq k<j\leq
n}r_{k}r_{j}cov\{e_{k},e_{j}\}=O(\log^{4}n/n)\,.
\]
We can therefore conclude that for any $\delta>0$
\[
\sum_{k=1}^{\infty}\frac{2^{k}}{k^{5}(\log k)^{1+\delta}}E[(\max_{n\leq2^{k}%
}|S_{n}|)^{2}]\leq\sum_{k=1}^{\infty}\frac{2^{k}}{k^{5}(\log k)^{1+\delta}%
}O\left(  \frac{k^{4}}{2^{k}}\right)  <\infty\,,
\]
since $\sum_{k=1}^{\infty}1/k(\log k)^{1+\delta}<\infty$, which by the
Borel-Cantelli lemma implies $\max_{n\leq2^{k}}|S_{n}|=o(k^{5/2}(\log
k)^{(1+\delta)/2}2^{-k/2})$ $a.s.$. Consequently $\sqrt{N}|S_{N}|=o((\log
N)^{5/2}(\log\log N)^{(1+\delta)/2})$ $a.s.$ since the function $(\log
n)^{5/2}(\log\log n)^{(1+\delta)/2}$ is slowly varying at infinity.

Now let $\widehat{d}_{T}$ denote the LPR-BA estimator. The analytically
bias-adjusted LPR estimator is the OLS coefficient of the regressor
$-2\log\lambda_{j}$ in the regression of $\log I_{T}(\lambda_{j})$ on $1$,
$-2\log\lambda_{j}$, and $\lambda_{j}^{2p}$, $p=1,\ldots,P $. Applying the
Frisch-Waugh-Lovell theorem and projecting out the regressors $\lambda
_{j}^{2p}$, $p=1,\ldots,P$, as well as unity we can express the estimation
error $\widehat{d}_{T}-d$ exactly as in \eqref{esterror}, save that the
$r_{j}$ are now defined in terms of $-2\widetilde{\log\lambda}_{j}$, say, the
component of $-2\log\lambda_{j}$ orthogonal to $1$ and $\lambda_{j}^{2p}$,
$p=1,\ldots,P$. This projection does not alter the overall magnitudes, so for
the orthogonalized regressor we have $\max_{1\leq j\leq N}|\widetilde
{\log\lambda}_{j}|=O(\log N)$ and $\sum_{j=1}^{N}(\widetilde{\log\lambda}%
_{j})^{2}=O(N)$ \citep[][Lemma 2,
parts (j) \& (k)]{andrews:guggenberger:2003}. The proof that $|\widehat{d}%
_{T}-d|\log T=o(1)$ $a.s.$ now proceeds as previously with $r_{j}%
=\widetilde{\log\lambda}_{j}/\sum_{j=1}^{N}(\widetilde{\log\lambda}_{j})^{2}$,
$j=1,\ldots,N$.

For the SPLW estimator the proposition follows directly from \citet[][Lemma
5.8]{giraitis:robinson:2003}, which implies that the SPLW estimator satisfies
$P(|\widehat{d}_{T}-d|\log T>\epsilon)=o(N^{-p})$, where $p>1/\epsilon$ and
$N$, the bandwidth, satisfies $T^{\epsilon}<N<T^{1-\epsilon}$ for some
$\epsilon>0$. For the SPLW-BA estimator the proposition can be established in
a manner similar to that employed above for the LPR and LPR-BA estimators.
Using Lemma 4 of \cite{andrew:sun:2004} we can express $\widehat{d}_{T}-d$,
where $\widehat{d}_{T}$ now denotes the SPLW-BA estimator, as a function of
the standardized score and from Lemma 5 of \cite{andrew:sun:2004} we can
conclude that the standardized score is of an order that implies that
$|\widehat{d}_{T}-d|\log T=o(1)$ $a.s.$, \textit{cf}.
\citet[][Theorem 4]{andrew:sun:2004}.\hfill\qed

\bigskip

\section{Tables}


\begin{landscape}

\begin{table}[tbp] \centering
\caption{Bias and mean square error (MSE) for all LPR-based estimators: $T=100.$
Unadjusted (LPR); analytically bias-adjusted (LPR-BA); bootstrap bias-adjusted
(LPR$_{sb}$ for $k=0,1,2$); bootstrap bias-adjusted \textit{after }analytical adjustment
(LPR-BA$_{sb}$). The lowest bias (in absolute value) and MSE for each
parameter setting are highlighted in bold.}\label{table_lpr_bias_100}%

\begin{tabular}
[c]{|c|c|c|c|c|c||c|c|c|c||c|c|c|c|}\hline\hline
&  &  & \multicolumn{3}{|c}{} & \multicolumn{4}{|c}{} &
\multicolumn{4}{|c|}{$LPR$-$BA_{sb}$}\\
&  &  & \multicolumn{3}{|c}{$LPR$-$BA$} & \multicolumn{4}{|c|}{$LPR_{sb}$} &
\multicolumn{3}{|c|}{$P=1$} & $P=2$\\
&  & $LPR$ & $P=1$ & $P=2$ & $P=3$ & $k=0$ & $k=1$ & $k=2$ & $SSR$ & $k=0$ &
$k=1$ & $SSR$ & $k=0$\\\hline
$d$ & $\phi$ & \multicolumn{12}{|c|}{Bias}\\\hline
0 & 0.3 & 0.1445 & 0.0366 & 0.0138 & 0.0236 & 0.1255 & 0.0944 & 0.0368 &
0.0798 & 0.0165 & \textbf{-0.0063} & 0.0108 & -0.0161\\
& 0.6 & 0.3947 & 0.2000 & 0.1039 & 0.0725 & 0.3511 & 0.2799 & 0.1506 &
0.2655 & 0.1574 & 0.0919 & 0.1485 & \textbf{0.0564}\\
& 0.9 & 0.8230 & 0.7402 & 0.6540 & \textbf{0.5969} & 0.8000 & 0.7188 &
0.8053 & 0.7439 & 0.7031 & 0.6234 & 0.6915 & 0.6161\\
0.2 & 0.3 & 0.1400 & 0.0395 & 0.0161 & 0.0262 & 0.1207 & 0.0886 & 0.0252 &
0.0769 & 0.0220 & -0.0116 & 0.0152 & \textbf{-0.0093}\\
& 0.6 & 0.3887 & 0.2017 & 0.1047 & 0.0746 & 0.3401 & 0.2609 & 0.1183 &
0.2459 & 0.1549 & 0.0805 & 0.1474 & \textbf{0.0601}\\
& 0.9 & 0.7968 & 0.7310 & 0.6558 & \textbf{0.5937} & 0.8180 & 0.8612 &
0.9425 & 0.8900 & 0.7301 & 0.6253 & 0.7162 & 0.6534\\
0.4 & 0.3 & 0.1374 & 0.0461 & 0.0194 & 0.0309 & 0.1110 & 0.0684 & -0.0130 &
0.0590 & 0.0229 & -0.0193 & 0.0127 & \textbf{-0.0047}\\
& 0.6 & 0.3780 & 0.2051 & 0.1063 & 0.0730 & 0.3319 & 0.2555 & 0.1178 &
0.2349 & 0.1546 & 0.0713 & 0.1368 & \textbf{0.0620}\\
& 0.9 & 0.7245 & 0.6910 & 0.6333 & \textbf{0.5706} & 0.8107 & 0.9839 &
1.2222 & 1.1407 & 0.7485 & 0.8018 & 0.7676 & 0.6859\\\hline
$d$ & $\phi$ & \multicolumn{12}{|c|}{MSE}\\\hline
0 & 0.3 & \textbf{0.0463} & 0.0753 & 0.1483 & 0.2369 & 0.0650 & 0.1396 &
0.3867 & 0.1869 & 0.1349 & 0.2549 & 0.1602 & 0.2525\\
& 0.6 & 0.1810 & \textbf{0.1125} & 0.1543 & 0.2348 & 0.1711 & 0.2041 &
0.4095 & 0.2461 & 0.1515 & 0.2807 & 0.1841 & 0.2483\\
& 0.9 & 0.7031 & 0.6189 & \textbf{0.5658} & 0.5861 & 0.6948 & 0.7188 &
0.8053 & 0.7552 & 0.6235 & 0.6697 & 0.6471 & 0.6341\\
0.2 & 0.3 & \textbf{0.0449} & 0.0737 & 0.1409 & 0.2247 & 0.0612 & 0.1276 &
0.3602 & 0.1664 & 0.1187 & 0.2478 & 0.1532 & 0.2220\\
& 0.6 & 0.1765 & \textbf{0.1117} & 0.1493 & 0.2310 & 0.1640 & 0.1942 &
0.4002 & 0.2357 & 0.1422 & 0.2664 & 0.1620 & 0.2301\\
& 0.9 & 0.6589 & 0.6026 & \textbf{0.5620} & 0.5752 & 0.7375 & 0.9677 &
1.6844 & 1.2435 & 0.6903 & 0.7542 & 0.7804 & 0.7180\\
0.4 & 0.3 & \textbf{0.0440} & 0.0747 & 0.1415 & 0.2372 & 0.0616 & 0.1201 &
0.3477 & 0.1507 & 0.1174 & 0.2565 & 0.1585 & 0.2253\\
& 0.6 & 0.1676 & \textbf{0.1135} & 0.1498 & 0.2403 & 0.1635 & 0.2106 &
0.4723 & 0.2811 & 0.1486 & 0.2976 & 0.1944 & 0.2420\\
& 0.9 & 0.5519 & 0.5458 & \textbf{0.5325} & 0.5532 & 0.7585 & 1.3521 &
2.8938 & 2.3084 & 0.7647 & 1.2723 & 1.0173 & 0.8160\\\hline\hline
\end{tabular}
\end{table}%

\begin{table}[tbp] \centering
\caption{Bias and mean square error (MSE) for all LPR-based estimators: $T=500.$
Unadjusted (LPR); analytically bias-adjusted (LPR-BA); bootstrap bias-adjusted
(LPR$_{sb}$ for $k=0,1,2$); bootstrap bias-adjusted \textit{after }analytical
adjustment (LPR-BA$_{sb}$). The lowest bias (in absolute value) and MSE for each
parameter setting are highlighted in bold.}\label{table_lpr_bias_500}%

\begin{tabular}
[c]{|c|c|c|c|c|c||c|c|c|c||c|c|c|c|}\hline\hline
&  &  & \multicolumn{3}{|c}{} & \multicolumn{4}{|c}{} &
\multicolumn{4}{|c|}{$LPR$-$BA_{sb}$}\\
&  &  & \multicolumn{3}{|c}{$LPR$-$BA$} & \multicolumn{4}{|c|}{$LPR_{sb}$} &
\multicolumn{3}{|c|}{$P=1$} & $P=2$\\
&  & $LPR$ & $P=1$ & $P=2$ & $P=3$ & $k=0$ & $k=1$ & $k=2$ & $SSR$ & $k=0$ &
$k=1$ & $SSR$ & $k=0$\\\hline
$d$ & $\phi$ & \multicolumn{12}{|c|}{Bias}\\\hline
0 & 0.3 & 0.0619 & 0.0097 & 0.0060 & 0.0026 & 0.0351 & -0.0020 & -0.0607 &
-0.0053 & 0.0025 & -0.0089 & 0.0018 & \textbf{0.0001}\\
& 0.6 & 0.2221 & 0.0671 & 0.0244 & 0.0090 & 0.1603 & 0.0652 & -0.1000 &
0.0446 & 0.0282 & -0.0323 & 0.0168 & \textbf{-0.0016}\\
& 0.9 & 0.6736 & 0.4946 & 0.3707 & 0.2814 & 0.5927 & 0.4628 & \textbf{0.2351}
& 0.4014 & 0.4114 & 0.2818 & 0.3802 & 0.2917\\
0.2 & 0.3 & 0.0601 & 0.0101 & 0.0066 & 0.0036 & 0.0330 & -0.0044 & -0.0642 &
-0.0063 & 0.0020 & -0.0108 & 0.0019 & \textbf{-0.0014}\\
& 0.6 & 0.2205 & 0.0679 & 0.0253 & 0.0105 & 0.1561 & 0.0570 & -0.1140 &
0.0344 & 0.0270 & -0.0353 & 0.0166 & \textbf{-0.0027}\\
& 0.9 & 0.6691 & 0.4948 & 0.3713 & 0.2840 & 0.5972 & 0.4758 & \textbf{0.2610}
& 0.4168 & 0.4045 & 0.2660 & 0.3765 & 0.2842\\
0.4 & 0.3 & 0.0613 & 0.0151 & 0.0126 & 0.0110 & 0.0320 & -0.0079 & -0.0720 &
-0.0087 & 0.0034 & -0.0126 & 0.0049 & \textbf{0.0000}\\
& 0.6 & 0.2206 & 0.0725 & 0.0304 & 0.0174 & 0.1488 & 0.0392 & -0.1489 &
0.0116 & 0.0262 & -0.0418 & 0.0190 & \textbf{-0.0041}\\
& 0.9 & 0.6534 & 0.4908 & 0.3704 & \textbf{0.2856} & 0.6621 & 0.6785 &
0.6175 & 0.6529 & 0.4227 & 0.3126 & 0.3958 & 0.2876\\\hline
$d$ & $\phi$ & \multicolumn{12}{|c|}{MSE}\\\hline
0 & 0.3 & \textbf{0.0103} & 0.0165 & 0.0293 & 0.0409 & 0.0131 & 0.0271 &
0.0783 & 0.0360 & 0.0236 & 0.0414 & 0.0284 & 0.0389\\
& 0.6 & 0.0558 & \textbf{0.0210} & 0.0302 & 0.0413 & 0.0385 & 0.0400 &
0.1237 & 0.0817 & 0.0288 & 0.0624 & 0.0554 & 0.0463\\
& 0.9 & 0.4603 & 0.2614 & 0.1675 & \textbf{0.1208} & 0.3636 & 0.2468 &
0.1611 & 0.3356 & 0.1999 & 0.1549 & 0.2361 & 0.1393\\
0.2 & 0.3 & \textbf{0.0102} & 0.0168 & 0.0303 & 0.0420 & 0.0130 & 0.0272 &
0.0782 & 0.0310 & 0.0235 & 0.0404 & 0.0307 & 0.0387\\
& 0.6 & 0.0552 & \textbf{0.0213} & 0.0307 & 0.0416 & 0.0371 & 0.0382 &
0.1253 & 0.0798 & 0.0288 & 0.0620 & 0.0511 & 0.0456\\
& 0.9 & 0.4542 & 0.2614 & 0.1675 & \textbf{0.1221} & 0.3715 & 0.2672 &
0.1980 & 0.3487 & 0.1941 & 0.1432 & 0.2369 & 0.1334\\
0.4 & 0.3 & \textbf{0.0103} & 0.0169 & 0.0303 & 0.0415 & 0.0127 & 0.0261 &
0.0748 & 0.0274 & 0.0219 & 0.0351 & 0.0265 & 0.0356\\
& 0.6 & 0.0552 & \textbf{0.0219} & 0.0312 & 0.0420 & 0.0345 & 0.0349 &
0.1288 & 0.0953 & 0.0278 & 0.0587 & 0.0446 & 0.0430\\
& 0.9 & 0.4342 & 0.2579 & 0.1678 & \textbf{0.1243} & 0.4631 & 0.5524 &
0.7172 & 0.5893 & 0.2180 & 0.2031 & 0.2479 & 0.1446\\\hline\hline
\end{tabular}
\end{table}%

\begin{table}[tbp] \centering
\caption{Empirical coverage and length of (nominal 95\%) HPD intervals for all LPR-based
estimators: $T=100,500.$ Unadjusted (LPR); analytically bias-adjusted
(LPR-BA); bootstrap bias-adjusted (LPR$_{sb}$ for $k=0,1,2$); bootstrap
bias-adjusted \textit{after }analytical adjustment (LPR-BA$_{sb}$). Figures
are averaged over all values of $d$ used in the experimental design for each
value of $\phi .$ Coverages for the intervals based on the asymptotic
distribution of the LPR and analytically bias-adjusted (LPR-BA) estimators
are also reported for comparison. The empirical coverage closest to
the nominal 95\%, and the shortest length, are highlighted in bold.}\label{table_lpr_HPD}%

\begin{tabular}
[c]{|c|c|c|c|c||c|c|c||c|c|c||c|c|c|}\hline\hline
&  &  & \multicolumn{2}{|c}{} & \multicolumn{3}{|c}{} &
\multicolumn{3}{|c|}{$LPR$-$BA_{sb}$} &
\multicolumn{3}{|c|}{\textit{Asymptotic interval}}\\
&  &  & \multicolumn{2}{|c|}{$LPR$-$BA$} & \multicolumn{3}{|c|}{$LPR_{sb}$} &
\multicolumn{2}{|c|}{$P=1$} & $P=2$ &  & \multicolumn{2}{|c|}{$LPR$-$BA$}\\
&  & $LPR$ & $P=1$ & $P=2$ & $k=0$ & $k=1$ & $k=2$ & $k=0$ & $k=1$ & $k=0$ &
$LPR$ & $P=1$ & $P=2$\\\hline
$\phi$ & $T$ & \multicolumn{12}{|c|}{Coverage}\\\hline
0.3 & 100 & 0.9015 & 0.9795 & 0.9730 & 0.9048 & 0.8880 & 0.8410 & 0.9773 &
\textbf{0.9612} & 0.9635 & {0.7563} & 0.8408 & 0.8035\\
& 500 & 0.8793 & 0.9748 & 0.9698 & 0.9120 & 0.9128 & 0.9075 & 0.9683 &
\textbf{0.9555} & 0.9703 & {0.8343} & 0.9083 & 0.8873\\\hline
0.6 & 100 & 0.2058 & 0.9160 & 0.9713 & 0.2475 & 0.3010 & 0.3328 & 0.9248 &
0.9092 & \textbf{0.9595} & {0.1918} & 0.7078 & 0.7860\\
& 500 & 0.0698 & 0.9388 & 0.9710 & 0.0898 & 0.1590 & 0.2155 & 0.9435 &
\textbf{0.9440} & 0.9738 & {0.1593} & 0.8565 & 0.8840\\\hline
0.9 & 100 & 0.0000 & 0.1568 & \textbf{0.5945} & 0.0005 & 0.0013 & 0.0195 &
0.1898 & 0.2405 & 0.5880 & {0.0010} & 0.1020 & 0.3065\\
& 500 & 0.0000 & 0.0030 & 0.2150 & 0.0000 & 0.0000 & 0.0005 & 0.0063 &
0.0140 & \textbf{0.3168} & {0.0000} & 0.0200 & 0.2670\\\hline
$\phi$ & $T$ & \multicolumn{12}{|c|}{Interval length}\\\hline
0.3 & 100 & \textbf{0.6413} & 1.1082 & 1.5664 & 0.6425 & 0.6452 & 0.6507 &
1.1070 & 1.0978 & 1.5542 & {0.5016} & 0.7523 & 0.9404\\
& 500 & 0.3278 & 0.5267 & 0.6976 & \textbf{0.3275} & 0.3279 & 0.3303 &
0.5271 & 0.5275 & 0.6984 & {0.2856} & 0.4283 & 0.5354\\\hline
0.6 & 100 & 0.6404 & 1.1046 & 1.5622 & 0.6409 & 0.6410 & \textbf{0.6392} &
1.1045 & 1.0931 & 1.5492 & {0.5016} & 0.7523 & 0.9404\\
& 500 & 0.3308 & 0.5274 & 0.6983 & 0.3294 & \textbf{0.3290} & 0.3306 &
0.5269 & 0.5273 & 0.6989 & {0.2856} & 0.4283 & 0.5354\\\hline
0.9 & 100 & 0.6114 & 1.0347 & 1.4638 & 0.6056 & 0.5716 & \textbf{0.5062} &
0.9663 & 0.9251 & 1.3499 & {0.5016} & 0.7523 & 0.9404\\
& 500 & 0.3306 & 0.5224 & 0.6954 & 0.3325 & 0.3252 & \textbf{0.3150} &
0.5241 & 0.5244 & 0.6957 & {0.2856} & 0.4283 & 0.5354\\\hline\hline
\end{tabular}
\end{table}%

\begin{table}[tbp] \centering
\caption{Bias and mean square error (MSE) for all SPLW-based estimators: $T=100.$
Unadjusted (SPLW); analytically bias-adjusted (SPLW-BA); bootstrap bias-adjusted
(SPLW$_{sb}$ for $k=0,1,2$); bootstrap bias-adjusted \textit{after }analytical adjustment
(SPLW-BA$_{sb}$). The lowest bias (in absolute value) and MSE for each
parameter setting are highlighted in bold.}\label{table_lpw_bias_100}%

\begin{tabular}
[c]{|c|c|c|c|c|c||c|c|c|c||c|c|c|c|}\hline\hline
&  &  & \multicolumn{3}{|c}{} & \multicolumn{4}{|c}{} &
\multicolumn{4}{|c|}{$SPLW$-$BA_{sb}$}\\
&  &  & \multicolumn{3}{|c}{$SPLW$-$BA$} & \multicolumn{4}{|c|}{$SPLW_{sb}$} &
\multicolumn{3}{|c|}{$P=1$} & $P=2$\\
&  & $SPLW$ & $P=1$ & $P=2$ & $P=3$ & $k=0$ & $k=1$ & $k=2$ & $SSR$ & $k=0$ &
$k=1$ & $SSR$ & $k=0$\\\hline
$d$ & $\phi$ & \multicolumn{12}{|c|}{Bias}\\\hline
0 & 0.3 & 0.1327 & \textbf{-0.0064} & -0.0393 & -0.0715 & 0.1191 & 0.0997 &
0.0647 & 0.1003 & 0.0111 & 0.0315 & 0.0078 & -0.0250\\
& 0.6 & 0.3993 & 0.1629 & 0.0530 & \textbf{-0.0214} & 0.3697 & 0.3243 &
0.2456 & 0.3252 & 0.1530 & 0.1328 & 0.1492 & 0.0504\\
& 0.9 & 0.8239 & 0.7139 & 0.6192 & \textbf{0.5165} & 0.8164 & 0.8035 &
0.7706 & 0.8043 & 0.7125 & 0.7044 & 0.7097 & 0.6209\\
0.2 & 0.3 & 0.1268 & \textbf{-0.0058} & -0.0397 & -0.0709 & 0.1127 & 0.0928 &
0.0572 & 0.0929 & 0.0119 & 0.0311 & 0.0084 & -0.0216\\
& 0.6 & 0.3922 & 0.1633 & 0.0538 & \textbf{-0.0214} & 0.3586 & 0.3062 &
0.2133 & 0.3072 & 0.1494 & 0.1228 & 0.1469 & 0.0535\\
& 0.9 & 0.7997 & 0.7029 & 0.6154 & \textbf{0.5068} & 0.8296 & 0.8687 &
0.8438 & 0.8755 & 0.7227 & 0.7113 & 0.7207 & 0.6378\\
0.4 & 0.3 & 0.1246 & \textbf{0.0004} & -0.0340 & -0.0668 & 0.1081 & 0.0842 &
0.0395 & 0.0846 & 0.0129 & 0.0234 & 0.0109 & -0.0141\\
& 0.6 & 0.3831 & 0.1668 & 0.0586 & \textbf{-0.0193} & 0.3534 & 0.3035 &
0.2060 & 0.3039 & 0.1466 & 0.1124 & 0.1460 & 0.0565\\
& 0.9 & 0.7363 & 0.6724 & 0.5942 & \textbf{0.4913} & 0.8291 & 0.8785 &
0.7524 & 0.9266 & 0.7419 & 0.6804 & 0.7288 & 0.6583\\\hline
$d$ & $\phi$ & \multicolumn{12}{|c|}{MSE}\\\hline
0 & 0.3 & \textbf{0.0352} & 0.0523 & 0.1128 & 0.1993 & 0.0393 & 0.0624 &
0.1518 & 0.0623 & 0.0830 & 0.1541 & 0.0896 & 0.1790\\
& 0.6 & 0.1787 & \textbf{0.0789} & 0.1129 & 0.1921 & 0.1621 & 0.1556 &
0.1999 & 0.1562 & 0.1008 & 0.1720 & 0.1070 & 0.1777\\
& 0.9 & 0.6969 & 0.5620 & 0.4913 & \textbf{0.4533} & 0.6973 & 0.7108 &
0.7566 & 0.7116 & 0.5869 & 0.6475 & 0.5835 & 0.5553\\
0.2 & 0.3 & \textbf{0.0339} & 0.0522 & 0.1104 & 0.1944 & 0.0379 & 0.0605 &
0.1479 & 0.0602 & 0.0766 & 0.1437 & 0.0832 & 0.1550\\
& 0.6 & 0.1732 & \textbf{0.0789} & 0.1105 & 0.1885 & 0.1551 & 0.1488 &
0.1973 & 0.1493 & 0.0975 & 0.1627 & 0.0992 & 0.1628\\
& 0.9 & 0.6575 & 0.5451 & 0.4853 & \textbf{0.4382} & 0.7311 & 0.8819 &
1.0841 & 0.8835 & 0.6199 & 0.7251 & 0.6557 & 0.6122\\
0.4 & 0.3 & \textbf{0.0334} & 0.0524 & 0.1088 & 0.1934 & 0.0368 & 0.0594 &
0.1454 & 0.0593 & 0.0739 & 0.1356 & 0.0728 & 0.1461\\
& 0.6 & 0.1660 & \textbf{0.0803} & 0.1116 & 0.1892 & 0.1565 & 0.1635 &
0.2385 & 0.1663 & 0.1000 & 0.1714 & 0.1011 & 0.1639\\
& 0.9 & 0.5619 & 0.5027 & 0.4597 & \textbf{0.4249} & 0.7508 & 1.1471 &
1.4360 & 1.0075 & 0.6790 & 0.8647 & 0.7115 & 0.6851\\\hline\hline
\end{tabular}%
\end{table}%

\begin{table}[tbp] \centering
\caption{Bias and mean square error (MSE) for all SPLW-based estimators: $T=500.$
Unadjusted (SPLW); analytically bias-adjusted (SPLW-BA); bootstrap bias-adjusted
(SPLW$_{sb}$ for $k=0,1,2$); bootstrap bias-adjusted \textit{after }analytical
adjustment (SPLW-BA$_{sb}$). The lowest bias (in absolute value) and MSE for each
parameter setting are highlighted in bold.}\label{table_lpw_bias_500}%

\begin{tabular}
[c]{|c|c|c|c|c|c||c|c|c|c||c|c|c|c|}\hline\hline
&  &  & \multicolumn{3}{|c}{} & \multicolumn{4}{|c}{} &
\multicolumn{4}{|c|}{$SPLW$-$BA_{sb}$}\\
&  &  & \multicolumn{3}{|c}{$SPLW$-$BA$} & \multicolumn{4}{|c|}{$SPLW_{sb}$} &
\multicolumn{3}{|c|}{$P=1$} & $P=2$\\
&  & $SPLW$ & $P=1$ & $P=2$ & $P=3$ & $k=0$ & $k=1$ & $k=2$ & $SSR$ & $k=0$ &
$k=1$ & $SSR$ & $k=0$\\\hline
$d$ & $\phi$ & \multicolumn{12}{|c|}{Bias}\\\hline
0 & 0.3 & 0.0573 & -0.0058 & -0.0130 & -0.0320 & 0.0323 & -0.0013 & -0.0517 &
-0.0009 & 0.0012 & 0.0076 & -0.0014 & \textbf{0.0000}\\
& 0.6 & 0.2306 & 0.0550 & 0.0068 & -0.0255 & 0.1755 & 0.0920 & -0.0501 &
0.0922 & 0.0286 & -0.0117 & 0.0293 & \textbf{0.0005}\\
& 0.9 & 0.7250 & 0.5273 & 0.3849 & \textbf{0.2659} & 0.6762 & 0.6045 &
0.4876 & 0.6050 & 0.4765 & 0.4002 & 0.4770 & 0.3340\\
0.2 & 0.3 & 0.0564 & -0.0038 & -0.0108 & -0.0293 & 0.0316 & -0.0018 &
-0.0513 & -0.0012 & 0.0030 & 0.0091 & \textbf{0.0000} & 0.0009\\
& 0.6 & 0.2292 & 0.0569 & 0.0090 & -0.0228 & 0.1716 & 0.0847 & -0.0630 &
0.0849 & 0.0292 & -0.0122 & 0.0302 & \textbf{0.0017}\\
& 0.9 & 0.7195 & 0.5269 & 0.3854 & \textbf{0.2679} & 0.6846 & 0.6298 &
0.5374 & 0.6304 & 0.4696 & 0.3839 & 0.4685 & 0.3265\\
0.4 & 0.3 & 0.0582 & 0.0018 & -0.0046 & -0.0227 & 0.0316 & -0.0040 & -0.0567 &
-0.0034 & 0.0048 & 0.0070 & 0.0034 & 0.0024\\
& 0.6 & 0.2296 & 0.0621 & 0.0146 & -0.0163 & 0.1664 & 0.0719 & -0.0889 &
0.0721 & 0.0292 & -0.0180 & 0.0300 & \textbf{0.0017}\\
& 0.9 & 0.7020 & 0.5222 & 0.3839 & \textbf{0.2697} & 0.7464 & 0.8296 &
0.8771 & 0.8312 & 0.4852 & 0.4267 & 0.4826 & 0.3283\\\hline
$d$ & $\phi$ & \multicolumn{12}{|c|}{MSE}\\\hline
0 & 0.3 & \textbf{0.0075} & 0.0106 & 0.0194 & 0.0300 & 0.0076 & 0.0134 &
0.0356 & 0.0135 & 0.0137 & 0.0202 & 0.0133 & 0.0231\\
& 0.6 & 0.0578 & \textbf{0.0137} & 0.0194 & 0.0295 & 0.0373 & 0.0241 &
0.0523 & 0.0241 & 0.0180 & 0.0348 & 0.0181 & 0.0281\\
& 0.9 & 0.5312 & 0.2907 & 0.1694 & \textbf{0.1010} & 0.4652 & 0.3801 &
0.2747 & 0.3809 & 0.2461 & 0.1984 & 0.2477 & 0.1453\\
0.2 & 0.3 & \textbf{0.0074} & 0.0106 & 0.0196 & 0.0302 & 0.0075 & 0.0131 &
0.0345 & 0.0132 & 0.0134 & 0.0193 & 0.0129 & 0.0226\\
& 0.6 & 0.0571 & \textbf{0.0139} & 0.0196 & 0.0298 & 0.0358 & 0.0223 &
0.0524 & 0.0224 & 0.0175 & 0.0332 & 0.0177 & 0.0276\\
& 0.9 & 0.5232 & 0.2903 & 0.1700 & \textbf{0.1024} & 0.4792 & 0.4201 &
0.3502 & 0.4209 & 0.2405 & 0.1878 & 0.2469 & 0.1403\\
0.4 & 0.3 & 0.0077 & 0.0108 & 0.0201 & 0.0305 & \textbf{0.0075} & 0.0130 &
0.0344 & 0.0131 & 0.0131 & 0.0181 & 0.0128 & 0.0221\\
& 0.6 & 0.0573 & \textbf{0.0147} & 0.0204 & 0.0302 & 0.0341 & 0.0205 &
0.0570 & 0.0205 & 0.0173 & 0.0322 & 0.0175 & 0.0273\\
& 0.9 & 0.4986 & 0.2854 & 0.1692 & \textbf{0.1037} & 0.5749 & 0.7557 &
1.1358 & 0.7496 & 0.2618 & 0.2452 & 0.2659 & 0.1461\\\hline\hline
\end{tabular}%
\end{table}%

\begin{table}[tbp] \centering
\caption{Empirical coverage and length of (nominal 95\%) HPD intervals for all SPLW-based
estimators: $T=100,500.$ Unadjusted (SPLW); analytically bias-adjusted
(SPLW-BA); bootstrap bias-adjusted (SPLW$_{sb}$ for $k=0,1,2$); bootstrap
bias-adjusted \textit{after }analytical adjustment (SPLW-BA$_{sb}$). Figures
are averaged over all values of $d$ used in the experimental design for each
value of $\phi .$ Coverages for the intervals based on the asymptotic
distribution of the SPLW and analytically bias-adjusted (SPLW-BA) estimators
are also reported for comparison. The empirical coverage closest to
the nominal 95\%, and the shortest length, are highlighted in bold.}\label{table_lpw_HPD}%

\begin{tabular}
[c]{|c|c|c|c|c||c|c|c||c|c|c||c|c|c|}\hline\hline
&  &  & \multicolumn{2}{|c}{} & \multicolumn{3}{|c}{} &
\multicolumn{3}{|c}{$SPLW$-$BA_{sb}$} &
\multicolumn{3}{|c|}{\textit{Asymptotic interval}}\\
&  &  & \multicolumn{2}{|c|}{$SPLW$-$BA$} & \multicolumn{3}{|c|}{$SPLW_{sb}$}
& \multicolumn{2}{|c|}{$P=1$} & $P=2$ &  & \multicolumn{2}{|c|}{$SPLW$-$BA$}\\
&  & $SPLW$ & $P=1$ & $P=2$ & $k=0$ & $k=1$ & $k=2$ & $k=0$ & $k=1$ & $k=0$ &
$SPLW$ & $P=1$ & $P=2$\\\hline
$\phi$ & $T$ & \multicolumn{12}{|c|}{Coverage}\\\hline
0.3 & 100 & 0.8563 & 0.9715 & 0.9643 & 0.8673 & 0.8773 & 0.8580 & 0.9718 &
\textbf{0.9630} & 0.9663 & {0.6765} & 0.7940 & 0.7565\\
& 500 & 0.7883 & 0.9685 & 0.9658 & 0.8645 & 0.8938 & 0.9010 & 0.9590 &
\textbf{0.9448} & 0.9648 & {0.7890} & 0.8978 & 0.8575\\\hline
0.6 & 100 & 0.1400 & 0.9268 & 0.9663 & 0.1503 & 0.1768 & 0.2000 & 0.9205 &
0.8955 & \textbf{0.9600} & {0.0713} & 0.6768 & 0.7480\\
& 500 & 0.0438 & 0.9300 & 0.9725 & 0.0480 & 0.0613 & 0.0725 & 0.9375 &
\textbf{0.9385} & 0.9663 & {0.0468} & 0.8505 & 0.8640\\\hline
0.9 & 100 & 0.0000 & 0.1268 & \textbf{0.5883} & 0.0000 & 0.0000 & 0.0018 &
0.1370 & 0.1549 & 0.5608 & {0.0000} & 0.0400 & 0.2205\\
& 500 & 0.0000 & 0.0020 & 0.1018 & 0.0000 & 0.0000 & 0.0000 & 0.0023 &
0.0045 & \textbf{0.1540} & {0.0000} & 0.0013 & 0.1100\\\hline
$\phi$ & $T$ & \multicolumn{12}{|c|}{Interval length}\\\hline
0.3 & 100 & \textbf{0.5400} & 0.9555 & 1.3830 & 0.5413 & 0.5445 & 0.5529 &
0.9569 & 0.9587 & 1.3789 & {0.3911} & 0.5866 & 0.7332\\
& 500 & \textbf{0.2630} & 0.4289 & 0.5770 & 0.2634 & 0.2645 & 0.2674 &
0.4291 & 0.4297 & 0.5775 & {0.2226} & 0.3340 & 0.4175\\\hline
0.6 & 100 & \textbf{0.5434} & 0.9562 & 1.3823 & 0.5445 & 0.5480 & 0.5538 &
0.9586 & 0.9592 & 1.3756 & {0.3911} & 0.5866 & 0.7332\\
& 500 & \textbf{0.2676} & 0.4305 & 0.5774 & 0.2680 & 0.2712 & 0.2809 &
0.4313 & 0.4340 & 0.5777 & {0.2226} & 0.3340 & 0.4175\\\hline
0.9 & 100 & 0.5034 & 0.8847 & 1.2921 & 0.4742 & 0.4248 & \textbf{0.4073} &
0.8236 & 0.7900 & 1.1904 & {0.3911} & 0.5866 & 0.7332\\
& 500 & 0.2638 & 0.4235 & 0.5753 & 0.2690 & 0.2632 & \textbf{0.2503} &
0.4270 & 0.4319 & 0.5785 & {0.2226} & 0.3340 & 0.4175\\\hline\hline
\end{tabular}%
\end{table}%

\end{landscape}%

\clearpage


\phantomsection
\addcontentsline{toc}{section}{References}
\bibliographystyle{ims}
\bibliography{tsa}


\end{document}